\documentclass[12pts]{article}

\usepackage{amsmath,amsthm, amssymb,epsfig,epsf,psfrag, color,soul,cancel}

\newtheorem{theorem}{Theorem}[section]

\newtheorem{corollary}{Corollary}[section]

\newcommand{\Prob} {{\bf P}}

\newcommand{\E}{{\bf E}}

\def\QED{{\setlength{\fboxsep}{0pt}\setlength{\fboxrule}
{0.2pt}\fbox{\rule[0pt]{0pt}{1.3ex}\rule[0pt]{1.3ex}{0pt}}}}

\begin{document}

% "Title of the paper"
\title{Scalable SUM-Shrinkage Schemes for Distributed Monitoring Large-Scale Data Streams}
\author{Kun Liu, Ruizhi Zhang \& Yajun Mei \\ % {\tt \{kliu,ymei\}@isye.gatech.edu} \\
H. Milton Stewart School of Industrial and Systems Engineering
\\ Georgia Institute of Technology \\
Atlanta, GA 30329-0205, USA}
%\date{August 28, 2009}
%\date{March 10, 2010}
%\date{May 12, 2010}
%\date{October 28, 2011}
%\date{February 28, 2015}
\date{September 9, 2015}
\maketitle

\begin{abstract}
%In a remarkable series of papers beginning in 1956, Charles Stein surprised
%the statistical community that the usual estimation of the normal mean vector
%could be dominated in dimensions $3$ and higher, and since then
%Shrinkage estimation has become a basic tool in the analysis of high-dimensional data,
%especially when the object to estimate holds sparsity properties.

In this article, motivated by biosurveillance and censoring sensor networks,
we investigate the problem of distributed monitoring large-scale
data streams where an undesired event may occur at some unknown
time and  affect only a few   unknown data streams. We propose to
develop scalable global monitoring schemes by  parallel running local detection
procedures and by combining these local procedures together to make a global decision
based on  SUM-shrinkage techniques. Our approach is illustrated in two concrete examples:
one is the nonhomogeneous case when the pre-change and post-change local distributions are given,
and the other is the homogeneous case of monitoring a large number of independent $N(0,1)$ data streams
where the means of some data streams might shift to unknown positive or negative values.
Numerical simulation  studies demonstrate the usefulness of the proposed schemes.
\end{abstract}

%The asymptotic properties of the average run length to false alarm of the proposed schemes are
%derived under the setting when the number of data streams goes to $\infty.$
%
%The sequential change-point detection problem is studied in a  general
%context of monitoring a large number of data streams when the ``trigger event" may affect different
%data streams differently, e.g., it could have an immediate or
%delayed impact on some unknown data streams.
%Motivated by the applications in censoring sensor networks and by the scenario when one has a prior knowledge
%that at most $r$ data streams will be affected, we propose a class of scalable global monitoring schemes
%based on the sum of those local CUSUM statistics that are ``large" under the order thresholding rule.
%Asymptotic analysis and numerical simulation  studies demonstrates the usefulness of the proposed schemes.


%==================================================================
\section{Introduction}

In the modern information age, one often faces the need to online monitor large-scale data streams with the aim of offering the potential for early detection of a ``trigger" event. Ideally, one would like to develop a global
monitoring scheme that can detect the occurring event as quickly as possible
while controlling the system-wise global false alarm rate.   From the statistical point of view,
this is a  sequential change-point detection or  quickest change detection  problem,
which has a variety of applications such as industrial quality control, signal detection and
biosurveillance. The classical version of this problem,
where one monitors independent and identically distributed (iid) {\it univariate} or {\it low-dimensional multivariate} observations
from  a single data stream,
is a well-developed area, and many classical procedures have been developed such as the Shewhart's chart (Shewhart \cite{shewhart:1931}),
moving average control charts, Page's CUSUM procedure (Page \cite{page:1954}), Shiryaev-Roberts procedure
(Shiryaev \cite{shiryayev:1963}, Roberts  \cite{roberts:1966}), window-limited procedures (Lai \cite{lai:1995}) and scan statistics (Glaz, Naus and Wallenstein \cite{scan}). All these classical procedures
not only hold attractive theoretical properties, but also are computationally simple.
See, for example, Lorden \cite{lorden:1971}, Pollak \cite{pollak:1985, pollak:1987}, Moustakides \cite{moustakides:1986},
Lai \cite{lai:1995, lai:2001}, Kulldorff \cite{kulldorff:2001}.  For a review,  see the books such as Basseville and  Nikiforov \cite{bass:1993}, Poor and Hadjiliadis \cite{poor2009quickest}, Tartakovsky, Nikiforov, and Basseville \cite{tartakovsky2014sequential}.

However, research is limited in the context of monitoring large-scale data streams, especially when the occurring event might
affect some, but not all, local data streams. The only exception is probably Xie and Siegmund \cite{xie2013sequential}, but their proposed schemes are computationally heavy with large local memory requirements to store past information, and thus is computationally infeasible for online monitoring large-scale data streams over long time period. Indeed, while many classical likelihood-ratio-based quickest change detection methods can be extended from  one or low dimension to high-dimension or large-scale data streams, they are generally computationally infeasible
in the context of large-scale data streams. As mentioned in Breiman \cite{breiman:2001}, in order for the profession of statistics to remain healthy, more algorithm-based methods should be developed. This is exactly what needs to be done in the subfield of quickest change detection or sequential change-point detection. We feel that the current main bottleneck is  on the algorithm or methodology aspect, and in particular, new ideas and new approaches are needed to develop efficient {\it scalable} global schemes in the sense of being able to be implemented for monitoring large-scale data streams over a long period of time.

\begin{figure}[t]\label{fig:mei1}
\centerline{
\includegraphics*[width=3.0in]{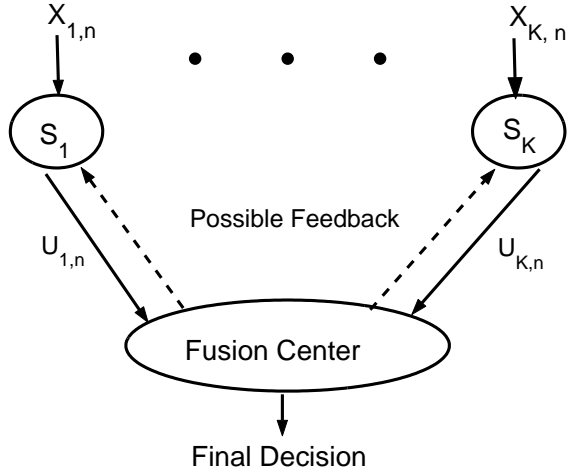}
} \caption{General setting of a widely used configuration of censoring
sensor networks.}
\end{figure}

The purpose of this article is to present a general and flexible approach that can provide efficient scalable global schemes when monitoring large-scale  data streams.
Our research is motivated by parallel and distributed computing and networks. A motivating example is censoring sensor networks in engineering, which was introduced by Rago, Willett, and Bar-Shalom \cite{rago:1996} and later by Appadwedula, Veeravalli, and Jones  \cite{appadwedula:2005} and Tay, Tsitsiklis, and Win \cite{tay:2007}.
Figure \ref{fig:mei1} illustrates the general setting of a widely used configuration of censoring sensor networks, in which the data streams $X_{k,n}$'s are observed at
the remote, distributed sensors, but the final decision is made at a central location, called the fusion center. The key feature of such a network is that while sensing (i.e., taking observations at the local sensors)  are generally cheap and affordable, communication between remote sensors and fusion center is expensive in terms of both energy and limited bandwidth. The question then becomes how the fusion center can still
monitor the system effectively  under the networks resource constraints.  A more concrete example is the National Syndromic Surveillance Program BioSense Platform at the Centers for Disease Control and Prevention (CDC), where the computing power and memory of any centralized server would have become limited as compared to {\it daily} summary data from all state and local health departments as well as many hospitals, and thus the CDC's BioSense Platform is designed to be a distributed computing system that can make a global decision.

To develop scalable schemes for distributed monitoring large-scale data streams, we propose to take advantage of parallel computing and the fact that many efficient and computationally simple local procedures are available to detect changes in local data streams. To be more specific, suppose we are monitoring a large number $K$  of data streams, and for each local data stream, we can construct a local detection procedure based upon some local detection statistics that can be computed recursively over time $n$, e.g., involving $O(1)$ computations and $O(1)$ memory requirements at each time. Then our proposed methodology is to run these $K$ local procedures in parallel before combining them into a global monitoring scheme. Hence it only requires $O(K)$ computations and $O(K)$ memory requirements at each time step when new observations are taken, thereby yielding a scalable  global monitoring scheme. While the parallel local monitoring approach sounds interesting, one allegation often made is that we will lose much information at the global level if we combine local detection procedures, not raw observation themselves, to make a global decision. Indeed, two specific methods have been developed in the literature to combine local detection procedures, but both have shortcomings: a naive method is to raise an alarm at the global level whenever any local detection procedure raises a local alarm, and the other method is developed in Mei \cite{mei:2009} to raise a global alarm when the sum of local detection statistics
is too large. Both methods are known to be inefficient when only  a few (unknown) subset of data streams are affected, see Mei \cite{mei:2009} and Xie and Siegmund \cite{xie2013sequential}.

In this article, we demonstrate that the problem is not on the parallel local monitoring approach itself, but on how to combine the local detection procedures suitably in the scenario when only a few (unknown) subset of data streams are affected. Our key idea is to filter out those unchanging local data streams and to make a global decision based on those likely affected data streams. For that purpose, we propose SUM-shrinkage techniques to combine the local detection statistics (in the log-likelihood ratio scale) of the local detection procedures together to make an efficient global decision.  It is worth pointing out that a well-known view in the standard off-line statistical inference literature is the necessity of shrinkage for high-dimensional data in order to improve power or efficiency. Thus, from the methodology point of view, our proposed methodologies are analogous to those off-line statistical methods such as (adaptive) truncation, and soft- and hard- thresholding,
see Neyman \cite{neyman:37}, Donoho and Johnstone \cite{donoho:94}, Fan and Lin \cite{Fan:98}.
Also see  Cand\`es \cite{candes:2006} and the references there. However, our motivation here is different and
our application to quickest change detection or sequential change-point detection is new.

The remainder of this article is organized as follows. In Section
2, we present our proposed ``SUM-shrinkage" methodology under a general setting of  monitoring
large-scale independent data streams and also discuss two existing methodologies for  parallel local monitoring.
We exemplify our  methodology in two concrete examples: Section 3 considers the censoring sensor networks when
the local data streams may or may not be homogeneous but the pre-change and post-change distributions of local data streams are given,
and Section 4 investigates a more complicated scenario when the post-change distributions of local data streams involve unknown parameters. In both Sections 3 and 4, numerical Monte Carlo simulation studies are conducted to illustrate the performance of our proposed methods. Section 5 includes the proofs of Theorems \ref{theorem1} and \ref{theorem2} that justify the choices of tuning/censoring parameters in censoring sensor networks.

\section{Our Proposed Methodology}

Let us present our proposed methodology under a general setting, and two specific examples will be given in later sections. Assume there are $K$  data streams in a system.
%, see Figure \ref{fig:mei1}:
\begin{eqnarray} \label{eqn01}
\mbox{Data Stream $1:$} && X_{1,1}, X_{1,2}, \cdots \\
\mbox{Data Stream $2:$} && X_{2,1}, X_{2,2}, \cdots \cr
\ldots && \ldots  \cr
\mbox{Data Stream $K:$} && X_{K,1}, X_{K,2}, \cdots. \nonumber
\end{eqnarray}
Initially, the system is ``in control", but at some {\it unknown} time $\nu,$ an undesired event may occur and   affect a few unknown local data streams in the sense of changing the local distributions of the $X_{k,n}$'s.

Here we assume that the online monitoring is conducted under the {\it unstructured} environment in the sense that we do not make any assumptions to relate the occurring event to the local data streams,  see Tartakovsky et al. \cite{tartakovsky:2006}, Mei \cite{mei:2009} and Xie and Siegmund \cite{xie2013sequential}.  Also see L{\'e}vy-Leduc and Roueff \cite{levy:2009} for an application of the unstructured problem to anomaly detection in computer networks. In particular, we focus on the scenario  when the occurring event changes the local distributions of affected local data streams,  and we do not aim to detect changes on the correlation between different data streams.
Hence,  the data $X_{k,n}$'s will be assumed to be independent across different data streams, but can be flexible otherwise. For instance, the $X_{k,n}$'s may or may not be identically distributed  across different local data streams, can be dependent over time within each local data stream, and can be {\it univariate} or {\it low-dimensional multivariate}. We should mention that the assumption of the independence across different data streams is standard in the unstructured problem,  see Tartakovsky et al. \cite{tartakovsky:2006}, Mei \cite{mei:2009} and Xie and Siegmund \cite{xie2013sequential}. This is not as restrictive as one thought in many practical applications, as the $X_{k,n}$'s can be chosen as the residuals of some spatio-temporal models rather than the original raw data. In other words, one can first pre-process the data, build a  spatio-temporal baseline model when the system is in control, calculate the corresponding residuals,
and then monitor  the residuals $X_{k,n}$'s. For an illustration, see Xie, Huang and Willett \cite{xie:2013a}, and Liu, Mei and Shi \cite{liu:2015}, to monitor the residuals of dependent data in two real-world applications in solar flare and hot-forming process.

For the purpose of generalization, we do not specify which kind of local changes  these $K$ data streams may have. Instead we assume that there is a local detection statistic $W_{k,n}$ (in the log-likelihood scale) for the $k$-th local data stream at each time step $n$ that summarizes the evidence regarding a possible local change based on the first $n$ local observations $(X_{k,1}, \ldots, X_{k,n})$  for each $k=1, \ldots, K.$ It is important that the $W_{k,n}$'s not only can detect local changes quickly, depending on specific assumptions on the local changes and local data distribution model,  but also can be computed recursively. We should emphasize that it can be highly non-trivial to construct such $W_{k,n}$'s, especially when the local  post-change distributions involve unknown parameters or when missing data are present, see our example in Section 4 and  Liu, Mei and Shi \cite{liu:2015}.  Here we assume, for a moment, that such $W_{k,n}$'s have been constructed, and our focus is the parallel local monitoring method that combines these local detection statistics $W_{k,n}$'s together to make an efficient global decision.

Before presenting our proposed methodology, let us first review the definition of global false alarm rate and two existing methods for parallel local monitoring.
When monitoring $K$ independent data streams in (\ref{eqn01}), it is well-known in statistics that even if each local false alarm rate is well controlled, the global false alarm rate can be significant when the number $K$ of data streams is large. In the literature of sequential change-point detection, for a global monitoring scheme that raise an alarm at time $T,$ its global false alarm rate is often evaluated by $1/ \E^{(\infty)}(T),$ where $\E^{(\infty)}(T)$ is the expectation of $T$ when the system is ``in control," and is often called the average run length to false alarm. A standard global false alarm constraint is to require  a  global monitoring scheme with a stopping time $T$ satisfying
 \begin{eqnarray} \label{eq000002}
\E^{(\infty)}(T) \ge \gamma,
\end{eqnarray}
where $\gamma > 0$ is a pre-specified constant. The rigorous definition of the detection delay of the scheme $T$  will be postponed later in next section.

There are two existing methods for parallel local monitoring. The first one is to raise an alarm at the global level whenever any local detection procedures raises a local alarm. If we normalize the local detection statistics $W_{k,n}$'s, this can be rewritten as raising an alarm at the global level at time
\begin{eqnarray} \label{eq3n02}
T_{\max}(a) = \inf\{n \ge 1: \max_{1 \le k \le K} W_{k,n} \ge a\},
\end{eqnarray}
($=\infty$ if such $n$ does not exist) where $a > 0$ is a pre-specified constant. Below we will call the scheme in (\ref{eq3n02})
the ``MAX" scheme.  The second method is the ``SUM" scheme developed in Mei \cite{mei:2009} that is defined by the stopping time
\begin{eqnarray} \label{eq3n03}
T_{{\rm sum}}(a) = \inf\{n \ge 1: \sum_{k=1}^{K} W_{k,n} \ge a \},
\end{eqnarray}
($=\infty$ if such $n$ does not exist). As mentioned in Mei \cite{mei:2009},
the ``MAX" scheme $T_{\max}(a)$ in (\ref{eq3n02}) works
well when one or very few data streams are affected, whereas
the ``SUM" scheme $T_{\rm sum}(a)$ in (\ref{eq3n03}) works well only when many data streams are affected.
Here and below the threshold $a$ of a scheme $T(a)$ is a pre-specified constant so that the scheme $T(a)$ satisfies the false alarm constraint $\gamma$ in (\ref{eq000002}).

%\section{Our Proposed Methodology}

Now we are ready to present our proposed methodology under a general setting. We suggest to define the global monitoring statistic of the general ``SUM-shrinkage" form
\begin{eqnarray} \label{shrinkage}
G_{n} = \sum_{k=1}^{K} h_{k}(W_{k,n}),
\end{eqnarray}
where $h_{k}(\cdot) \ge 0$ are some suitable shrinkage transformation functions. Then our proposed SUM-shrinkage scheme raises a global alarm at the time
\begin{eqnarray} \label{shrinkage2}
N_{G}(a) = \inf\{n \ge 1: G_{n} \ge a \}.
\end{eqnarray}
Intuitively,  the shrinkage functions $h_{k}$'s in (\ref{shrinkage}) play the role of dimension reduction by automatically filtering out those non-changing local data streams and by focusing only on those local data streams that appear to be affected by the occurring event.  In addition, as an extension, the proposed SUM-shrinkage scheme $N_{G}(a)$ enjoys the nice properties of the SUM scheme $T_{\rm sum}(a)$ in (\ref{eq3n03}): it does not assume that all local data streams are affected by the occurring event simultaneously, and thus can be very useful  when there may be a time delay before the event affects different local data streams, or when different local data streams involve local changes at different time steps. Also see Mei \cite{mei:2009} and Xie and Siegmund \cite{xie2013sequential} for  more discussions.

%thus the proposed scheme $N_{G}(a)$ in (\ref{shrinkage2}) will be efficient in the scenario when a few unknown local data streams are affected by the occurring event.

Evidently a suitable choice of the $h_{k}$'s in the SUM-shrinkage monitoring statistic $G_{n}$ in (\ref{shrinkage}) will depend on the assumptions and contexts of applications.  In Sections 3 and 4 below we will demonstrate the following three shrinkage transformations of the form
\hspace*{-10cm}  \begin{eqnarray}
\hspace*{-10cm}
&\bullet&  \mbox{Hard-thresholding: } h(x) = x {\bf 1}\{x \ge b\} \mbox{ for some constant $b,$}  \label{hard} \\
&\bullet&  \mbox{Soft-thresholding: } h(x) =  \max\{x - b, 0 \} \mbox{ for some constant $b,$}  \label{soft}  \\
&\bullet&  \mbox{Order-thresholding: } h(x) = x {\bf 1}\{x \ge w_{(r)}\}, \mbox{ where $w_{(r)}$ is the $r$-th largest statistic of $w_1, \cdots, w_{K}.$} \qquad \label{order}
\end{eqnarray}
The rationale and motivations of these three transformations will be given in Sections 3 and 4. Of course, besides those in (\ref{hard})-(\ref{order}),  there are many other kinds of the shrinkage functions such as $h(x) = \exp(b x).$ Also  by semi-Bayesian arguments, the transformation $h(x) = \log[1- p_0 + p_0 \exp(x)]$ is proposed and used in the schemes of Xie and Siegmund \cite{xie2013sequential} in a completely different manner under their setting.

We should emphasize that our proposed SUM-shrinkage schemes can be easily implemented in the distributed network systems as long as the local detection statistics $W_{k,n}$'s can be easily computed at local data streams. Besides the two examples in the next two sections, we should point out that our proposed ``SUM-shrinkage" methodology  in (\ref{shrinkage})-(\ref{shrinkage2}) has a broad range of other applications. For instance, the local detection statistics $W_{k,n}$'s can be defined for dependent observations such as those from the recursive schemes in Fuh and Mei \cite{Fuh:2015} for hidden Markov models, or those from the non-parametric detection schemes in Gorden and Pollak \cite{Gordon:1994}, depending on which kind of local models or local changes we are interested in. In addition, little information seems to be lost if we do not observe those local data streams with small values of $W_{k,n}$'s since they make limited contributions in our proposed global monitoring statistic $G_{n}$ in (\ref{shrinkage}). This motivated  Liu, Mei and Shi \cite{liu:2015}  to develop an efficient adaptive sensor relocation policy when one only has ability to observe $r$ out of $K$ data streams at each time step. This may occur in manufacturing process control when there are $K$ possible stages in the process but there are only $r$ expensive sensors available to monitor the process. In such a problem, the order-thresholding transformation can be combined with missing data techniques to be used not only in the global monitoring statistic $G_{n}$ in (\ref{shrinkage}) for quickest detection, but also in a greedy manner to adaptively observe those $r$ data streams with the largest $W_{k,n}$'s values at each time step. We feel the spirit of SUM-Shrinkage can have many other applications, and hopefully our research opens new research opportunities and directions, especially on monitoring large-scale data streams.

\subsection{General Guidelines}

Below we will provide some general guidelines on how to use our proposed SUM-shrinkage scheme $N_{G}(a)$ in (\ref{shrinkage2}). The performance of our proposed scheme will depend on two components: one is the local detection statistics $W_{k,n}$'s and the other is the shrinkage transformation functions $h_{k}$'s. The suitable choices of these two components depend on the applications and contexts, and will be demonstrated in the next two sections.

In general, the local detection statistics $W_{k,n}$'s should be able to efficiently detect local changes we are interested in, and ideally can also be recursively computed over time so that the local computation is simple.  Such a choice of the $W_{k,n}$'s might be straightforward in some applications (i.e., Section 3),  but can be highly non-trival in other cases (e.g., Section 4).  Also see Liu, Mei and Shi \cite{liu:2015} how to define $W_{k,n}$'s when the observations from some data streams are  missing or unobservable.

For the choice of the shrinkage transformation $h_{k}(\cdot)$'s, the situation can be complicated when the data streams are nonhomogeneous (or the $W_{k,n}$'s have different properties for different $k$'s). One rule of thumb is to choose $h_{k}(\cdot)$'s so that the local procedures based upon nonhomogeneous $W_{k,n}$'s will raise local alarms at roughly same time steps for all affected data streams, see our example in Section 3.

Given the choices of the local detection statistics $W_{k,n}$'s and the shrinkage transformation $h_{k}(\cdot)$'s, an important remaining question is how to determine the global threshold $a$ in (\ref{shrinkage2}) so that the proposed SUM-shrinkage scheme $N_{G}(a)$ in  (\ref{shrinkage2}) satisfies the global false alarm constraint $\gamma$ in (\ref{eq000002}). In general this is nontrivial, as it requires one to  accurately characterize the relationship between the threshold $a$ and the false alarm constraint $\gamma,$ when the dimension $K$ goes to $\infty.$ Intuitively, the global monitoring statistic $G_{n}$ in (\ref{shrinkage}) is the sum of $K$ (independent) random variables, one would expect that the central limited theorem (CLT) will be useful when the shrinkage transformation keeps most non-zero values, e.g., the hard-thresholding or soft-thresholding transformations in (\ref{hard}) or (\ref{soft}) when the censoring parameters $b$'s are not large, whereas the compound Poisson process will be needed when the shrinkage transformation only keeps very few non-zero values, e.g., the order-thresholding transformation in  (\ref{order}) with not so large $r$ value. The rigorous theoretical proofs are beyond the scope of this article and will be investigated elsewhere. Below we will use Chebyshev's inequality and CLT to provide two approximations of the global threshold $a$ in terms of $\gamma.$

To do so, let us assume that under the pre-change hypothesis $\Prob^{(\infty)},$  the local detection statistics $W_{k,n}$'s and their shrinkage transformations converge very fast to stationary distributions as the time step $n$ goes to $\infty.$ More specifically, we assume that for each $k,$ the $h_{k}(W_{k,n})$'s converge  to their limit $H_{k}^{*}$ which is stochastically larger than $h_{k}(W_{k,n})$'s and has a well-defined log-moment generating function
\[
\psi_{k}(\theta) = \log \E^{(\infty)}\exp(\theta H_{k}^{*})
\]
for all $\theta \ge 0.$ By the definition of $N_{G}(a)$ in (\ref{shrinkage2}) and by applying  Chebyshev's inequality to both $N_{G}(a) \ge 0$ and $\sum_{k=1}^{K} H_{k}^{*}$,
for any $x > 0,$
\begin{eqnarray*}
\E^{(\infty)}(N_{G}(a)) &\ge& x  \Prob^{(\infty)}\Big( N_{G}(a) \ge x \Big) \\
&=& x \Big[ 1 -  \Prob^{(\infty)}\Big( N_{G}(a) < x\Big) \Big] \\
&=& x \Big[ 1 -  \Prob^{(\infty)}\Big( \sum_{k=1}^{K} h_{k}(W_{k,n}) \ge a \mbox{ for some $1 \le n \le x$}\Big) \Big] \\
&\ge& x \Big[ 1 - x \Prob^{(\infty)}\Big( \sum_{k=1}^{K} H_{k}^{*} \ge a \Big)\Big] \\
&\ge& x  \Big[ 1 - x e^{-\theta a} \E^{(\infty)}\exp\Big(\theta \sum_{k=1}^{K} H_{k}^{*}\Big)\Big]  \\
&=& x \Big[ 1 - x e^{-\theta a} \exp\Big(\sum_{k=1}^{K} \psi_{k}(\theta)\Big) \Big],
%&=& x \Big[ 1 - x u_{\theta}(a)],
\end{eqnarray*}
where the last equation uses the fact that these $K$ data streams are independent across different data streams. Define
\[
u_{\theta}(a) = e^{-\theta a} \exp\Big(\sum_{k=1}^{K} \psi_{k}(\theta)\Big).
\]
Choosing $x$ to maximize $x (1 -x u_{\theta}(a)),$ we have $x = 1/(2 u_{\theta}(a)),$ and thus
$\E^{(\infty)}(N_{G}(a)) \ge 1 /(4 u_{\theta}(a))$
for all $a > 0$ and all $\theta > 0.$ Thus, % this suggests to choose $a$ satisfying $u_{\theta}(a) \le 1 /(4\gamma)$ for some $\theta > 0.$
in the special case when there exists a $\theta_0 > 0$ so that $\psi_{k}(\theta_0) \le 0$ for all $k,$ we have $u_{\theta}(a) \le \exp(-\theta_0 a)$ and
thus a simple choice of
\begin{eqnarray} \label{eqnheuris0}
a = (\log \gamma + \log 4) / \theta_0
\end{eqnarray}
will guarantee that $\E^{(\infty)}(N_{G}(a))  \ge \gamma.$ This implies that $a$ is of order $(\log \gamma) / \theta_0,$ where the $\theta_0$ value depends on the shrinkage transformation $h_{k}(\cdot).$
%In the sequential change-point detection literature, many local detection statistics $W_{k,n}$'s satisfy this condition with $\theta_0=1$ when they are based on the likelihood ratio tests.

Unfortunately, our numerical simulation suggests that the above choice of $a$ based on Chebyshev's inequalities is often too loose.
A better estimation of the global threshold $a$ can be found heuristically by using more refined approximations.
When the  CLT is applicable to the global monitoring statistic $G_{n}$ in (\ref{shrinkage}), we can use the approximation
\[
\Prob^{(\infty)}\Big( \sum_{k=1}^{K} H_{k}^{*} \ge a \Big) \approx \Prob\Big( N(0,1) \ge \frac{a - \mu_{H}}{\sigma_{H}}\Big),
\]
where $\mu_{H}$ and $\sigma_{H}^2$ are the mean and variance of the limiting global statistic  $\sum_{k=1}^{K} H_{k}^{*}:$
\[
\mu_{H} = \sum_{k=1}^{K} \dot \psi_{k}(0) \quad \mbox{ and } \quad \sigma_{H}^2 =  \sum_{k=1}^{K} \ddot \psi_{k}(0).
\]
In addition, if we approximate the distribution of $N_{G}(a)$ as an exponentially distribution, which is true to most sequential change-point detection schemes in the literature, then we have  $\E^{(\infty)}(N_{G}) \approx x / \Prob^{(\infty)}(N_{G}(a) \le x)$ for moderately large $x.$ Combining these above two approximations yields a heuristic approximation
\begin{eqnarray} \label{eqnheuris}
a = \mu_{H} + z_{1/\gamma} \sigma_{H},
\end{eqnarray}
where $z_{1/\gamma}=z$ so that $\Prob(N(0,1) \ge z) = 1 /\gamma.$ Our numerical simulation below supports this heuristic approximation, but rigorous justifications turn out to be highly technical due to the complicated correlation structures of the $W_{k,n}$'s over time domain $n,$ and will be presented elsewhere.

%but unfortunately we are unable to rigorously prove it so far, due to the complicated correlation structures of the $W_{k,n}$'s over time domain $n.$

\section{A First Example: % different shrinkage transformations $h_{k}$'s}
 %when
Censoring Sensor Networks with Known Post-Change Distributions}

For censoring sensor networks in Figure \ref{fig:mei1}, practitioners often prefer the local sensors to send summary messages $U_{k,n}$'s to the fusion center only when necessary, so as to prolong the reliability and lifetime of the network system. The question then becomes when and how to send summary messages so that the fusion center can still
monitor the network system effectively. In the quickest change detection problem in censoring sensor networks, let $X_{k,n}$ denote the observations at the $k$-th sensor at time  step $n.$
In this section, we focus on nonhomogeneous sensors, but make a restrictive assumption that the pre-change and post-change distributions of the $X_{k,n}$'s are given. In the next section, we will investigate the case of homogeneous sensors when the post-change distributions are unknown.

Specifically, in this section we follow the literature to assume that for each $k=1,\ldots, K,$ the density function of the observations at the $k$-th data stream is $f_{k}$ before the change, and is $g_{k}$ after the change if the $k$-th data stream is affected, where the $f_{k}$'s and $g_{k}$'s are completely specified densities with respect to a suitable measure $\mu$, see, for example, Tartakovsky and Veeravalli \cite{tartakovsky:2004}. For each $1 \le k
\le K,$ we assume that the Kullback-Leibler (KL) information number
\begin{eqnarray} \label{eq000001}
& & I(g_{k}, f_{k}) = \int \log\frac{g_{k}(x)}{f_{k}(x)} g_{k}(x)
d \mu(x)\
\end{eqnarray}
is finite and positive, and
\begin{eqnarray} \label{eq000001b}
\int \Big(\log\frac{g_{k}(x)}{f_{k}(x)} \Big)^2 g_{k}(x) d \mu(x)  <
\infty.
\end{eqnarray}

%From the statistical viewpoint, the assumptions of known $f_{k}$'s and known $g_{k}$'s are restrictive, but we want to emphasize that the corresponding problem is  non-trivial, especially when the sensors  are nonhomogeneous, as the fusion center may not have full access to raw observations.

%\medskip
\subsection{Our Proposed Methods}

Let us apply our proposed SUM-shrinkage schemes to censoring sensor networks in Figure $1.$ To do so, we need to define two components of our proposed global monitoring statistics in (\ref{shrinkage}). The first one is the local detection statistic $W_{k,n}$'s, which is simple in this context since the pre-change and post-change distributions, $f_{k}$ and $g_{k}$, are known at each local sensor. For instance, the $W_{k,n}$'s can be chosen as the well-known local CUSUM statistics  (Page \cite{page:1954}) that are defined recursively by
\begin{eqnarray} \label{eq3n01}
W_{k, n} = \max\Big( W_{k, n-1} + \log
\frac{g_{k}(X_{k,n})}{f_{k}(X_{k,n})},\ 0\Big),
\end{eqnarray}
for $n \ge 1$ and $W_{k,0}= 0$ for $k=1, \cdots, K.$ As shown in Lorden \cite{lorden:1971} and Moustakides \cite{moustakides:1986}, the local CUSUM statistics $W_{k,n}$'s in (\ref{eq3n01}) yield  the optimal local procedure to detect the local change under some suitable criteria.

Next, we need to specify concrete shrinkage transformation $h_{k}$'s in (\ref{shrinkage}) for censoring sensor networks. To prolong the reliability and lifetime of the network system, it is natural for the local sensors to transmit only those local CUSUM statistics $W_{k,n}$'s that are large. Specifically, at time $n,$  the sensor message from the sensor to the fusion center is given by
\begin{eqnarray} \label{hardthreshold}
U_{k, n} = \left\{%
\begin{array}{ll}
    W_{k,n}, & \hbox{if $W_{k,n} \ge b_{k}$} \\
    \mbox{NULL}, & \hbox{if $W_{k,n} < b_{k}$} \\\end{array}%
\right. ,
\end{eqnarray}
where $b_{k} \ge 0$ is the local censoring parameter at the $k$-th sensor
(or data stream). %Here the message ``NULL" is a special sensor symbol to indicate the local CUSUM statistic is not large.
In practice, the message ``NULL" could be represented by the situation when the
sensor does not send any message to the fusion center, e.g., the
sensor is silent.

After receiving the local sensor messages from the sensors, the fusion center then combines these local sensor messages $U_{k,n}$'s in (\ref{hardthreshold}) suitably together to make a global decision.  There are many approaches to do so, and below we illustrate three of them. The first two schemes are based on the summation of all sensor messages $U_{k,n}$'s, depending on how to interpret the ``NULL" values. If we treat the ``NULL" values as lower limit $0,$ then the fusion center raises a global alarm at time
\begin{eqnarray}
N_{hard}(a) &=& \inf\Big\{n \ge 1: \sum_{k=1}^{K} U_{k,n} \ge a\Big\} \cr
 &=& \inf\Big\{n \ge 1: \sum_{k=1}^{K} W_{k,n} {\bf 1}\{ W_{k,n} \ge b_{k}\} \ge a\Big\}. \label{eq000012}
\end{eqnarray}
Below this scheme will be referred as the hard-thresholding scheme,  since it is a special case of the global statistic in (\ref{shrinkage}) when the shrinkage functions $h_{k}$'s are  the hard-thresholding transformation in (\ref{hard}).

Meanwhile, if we treat the ``NULL" values as the upper limit $b_{k}$'s,  then the fusion center will  compute the global monitoring statistic
\[
G_{n} = \sum_{k=1}^{K} U_{k,n}= \sum_{k=1}^{K} \max\{ W_{k,n}, b_{k}\} = \sum_{k=1}^{K} \max\{ W_{k,n} - b_{k}, 0\} +  \sum_{k=1}^{K} b_{k},
\]
which is closely related to the soft-thresholding transformation in (\ref{soft}). Hence, we can define the soft-thresholding scheme that raises an alarm at time
\begin{eqnarray} \label{eq000012b}
N_{soft}(a) &=& \inf\Big\{n \ge 1: \sum_{k=1}^{K} \max\{ W_{k,n} - b_{k}, 0\} \ge a \Big\}.
\end{eqnarray}
Here we keep the threshold of $N_{soft}(a)$ as $a$ instead of $a - \sum_{k=1}^{K} b_{k},$ so that  $N_{soft}(a)$ is the special case of our proposed SUM-shrinkage scheme  $N_{G}(a)$ in (\ref{shrinkage2}) with the soft-thresholding transformation in (\ref{soft}).
%As a comparison, the performance of $N_{soft}(a)$ is comparable to that of $N_{hard}(a + \sum_{k=1}^{K} b_{k}).$

The third approach occurs when the fusion center has a prior knowledge that (at most) $r$ out of $K$ data streams
will be affected by the occurring event. Such a prior knowledge may be defined by the network fault-tolerant
design to avoid risking failure. In this case, it is reasonable for the fusion center to order all sensor messages $U_{k,n}$'s as $U_{(1), n} \ge \ldots \ge U_{(K),n},$ and raise an alarm if the sum of the $r$ largest $U_{k,n}$'s is too large. This is a combination of  the hard-thresholding transformation in (\ref{hard}) and the order-thresholding transformation in (\ref{order}), and it yields a global scheme that is defined by the stopping time
\begin{eqnarray} \label{rankCUSUM1}
N_{comb,r}(a) = \inf\Big\{n \ge 1: \sum_{k=1}^{r} U_{(k),n} \ge a \Big\}.
\end{eqnarray}
For simplicity, the ``NULL" values of $U_{k,n}$'s in the scheme $N_{comb,r}(a)$ in (\ref{rankCUSUM1})  will be treated  as the lower limit $0$ in our simulation below.

For the purpose of comparison, we  also apply the order-thresholding transformation in (\ref{order}) directly to the local CUSUM statistics $W_{k,n}$'s in (\ref{eq3n01}) themselves. Specifically, we
order the $K$ local CUSUM statistics $W_{1,n}, \ldots, W_{K,n}$ from largest to smallest:
$W_{(1), n} \ge W_{(2), n} \ge \ldots \ge W_{(K),n}.$ Then the order-thresholding  scheme can be defined by the stopping time
\begin{eqnarray} \label{rankCUSUM}
N_{order,r}(a) = \inf\Big\{n \ge 1: \sum_{k=1}^{r} W_{(k),n} \ge a \Big\}.
\end{eqnarray}
Of course, $N_{order,r}(a)$ is a special case of $N_{comb,r}(a)$ when the local censoring parameter $b_{k} \equiv 0,$ since the local CUSUM statistics $W_{k,n} \ge 0$ for all $k$  and all $n.$

% $N_{comb,r}(a)$ in (\ref{rankCUSUM1}) is a combination of the hard-thresholding scheme $N_{hard}(a)$ in (\ref{eq000012}) and the order-thresholding scheme $N_{order, r}(a)$ in (\ref{rankCUSUM}), whereas $N_{order,r}(a)$ is a special case of $N_{comb,r}(a)$ when $b_{k} \equiv 0.$ In addition, for  both $N_{comb,r}(a)$ and $N_{order, r}(a),$ the global statistic is also  of the form in (\ref{shrinkage}) when the shrinkage functions $h_{k}$'s are data-driven and adaptive over time.

%It is also interesting to mention that the order-thresholding scheme can also be thought of hard-thresholding schemes where the threshold parameters are data-driven and adaptive over time $n$. For instance, let $w_{r,n}$ denote the $r$-th largest statistics, i.e., $w_{r,n} = W_{(r), n},$ and then one raises an alarm at the global level at the time
%\begin{eqnarray*} %\label{rankCUSUMa}
%N_{order, r}^{*}(a) = \inf\Big\{n \ge 1: \sum_{k=1}^{K} W_{k,n} I\{W_{k,n} \ge w_{r,n}\} \ge a \Big\}.
%\end{eqnarray*}
%Rigorously speaking, we have $N_{order, r}(a) \le N_{order, r}^{*}(a),$ and they are
%equivalent only when the $W_{k,n}$'s are non-arithmetic, since they can be different when
%more than one of $W_{k,n}$'s is equal to $w_{r,n}.$ However, both $N_{order, r}(a)$ and $N_{order, r}^{*}(a)$
%possess similar asymptotic optimality properties and either can be used in our context.

\medskip
It is useful to mention that each of all four schemes in (\ref{eq000012})-(\ref{rankCUSUM}) is based on our proposed shrinkage statistics in (\ref{shrinkage}), but each  is actually a very large family of schemes that includes ``MAX"  or ``SUM" schemes or both as special cases.
For instance, for the hard-thresholding scheme $N_{hard}(a)$ in (\ref{eq000012}), it becomes  the ``SUM" scheme  $T_{\rm sum}(a)$ in (\ref{eq3n03}) if the censoring parameter $b_{k} \equiv 0$ for all $k,$ but becomes the ``MAX" scheme $T_{\rm max}(a)$ in (\ref{eq3n02})  if $b_{k} \equiv a$ for all $k.$ Similarly, the order-thresholding  scheme $N_{order,r}(a)$ in (\ref{rankCUSUM}) becomes the ``MAX" scheme when the order parameter $r=1$ and becomes  the ``SUM" scheme when $r = K.$ As for the soft-thresholding scheme $N_{soft}(a)$ in (\ref{eq000012b}),
it becomes  the ``SUM" scheme  if $b_{k} \equiv 0$ for all $k,$ and based on our numerical experience, its properties are similar to those of the ``MAX" scheme when $b_{k}$'s are very large.

Besides the local CUSUM statistics, another popular local detection statistic is the local Shiryaev-Roberts statistic (Shiryaev \cite{shiryayev:1963}, Roberts  \cite{roberts:1966}) which can be defined  in the log-likelihood ratio scale by
\begin{eqnarray} \label{eqn:SR}
{\hat W}_{k, n} = \log\Big( \exp({\hat W}_{k, n-1}) + 1 \Big) +  \log \frac{g_{k}(X_{k,n})}{f_{k}(X_{k,n})}
\end{eqnarray}
for $n \ge 1$ and  ${\hat W}_{k,0}= 0.$ It is well-known that the local Shiryaev-Roberts statistics ${\hat W}_{k, n}$  in (\ref{eqn:SR}) yield an efficient local detection procedure whose performance is similar to that of  local CUSUM statistics in (\ref{eq3n01}) when detecting a local change in distribution from $f_k$ to $g_k,$ see Pollak \cite{pollak:1985, pollak:1987}. In our numerical analysis below, the local detection statistics $W_{k,n}$'s can also  be defined as ${\hat W}_{k, n}$'s in (\ref{eqn:SR}), the local Shiryaev-Roberts statistics in logarithm scale,  or better yet, its positive part $\max\{{\hat W}_{k, n}, 0\}$. Our numerical simulation experiences suggest that the  performances of global monitoring schemes based upon local Shiryaev-Roberts statistics are similar to those based upon local CUSUM statistics in (\ref{eq3n01}) when monitoring $K$ data streams. Unfortunately  it is still an open question to investigate the theoretical properties of Shiryaev-Roberts-type schemes in the context of $K$ data streams, and thus we will focus on the local CUSUM statistics $W_{k, n}$'s in  (\ref{eq3n01}) as the local detection statistics below.

\subsection{Choices of the Thresholds $b_{k}$'s}

So far we simply follow our intuition without discussing how to
choose the local censoring parameters $b_{k}$'s in (\ref{hardthreshold}) for censoring sensor networks. Intuitively, the $b_{k}$'s should be
the same when the sensors are homogeneous, but they probably should be different when
the sensors are nonhomogeneous. It turns out that a ``good" choice
is
\begin{eqnarray} \label{eq000010}
b_{k} = \rho_{k} b
\end{eqnarray}
for $k=1, \ldots, K$ for some common constant $b \ge 0,$ where
\begin{eqnarray} \label{eq000011}
\rho_{k} = \frac{I(g_{k}, f_{k})}{\sum_{k=1}^{K} I(g_{k}, f_{k})}
\end{eqnarray}
and $I(g_{k}, f_{k})$ is the KL information number defined in (\ref{eq000001}).
Theoretical justification of our choice of $b_{k}$ in (\ref{eq000010})-(\ref{eq000011}) will be postponed to subsection 3.4.
Roughly speaking, $\rho_{k}$ in (\ref{eq000011}) can be thought of as the weight of the $k$-th data stream in the
overall final decision, and  the choice of $b_{k} = \rho_{k} b$ in (\ref{eq000010}) allows those affected local sensors
to send local messages $U_{k,n}$'s with large values to the fusion center at roughly the same time, thereby leading the quick detection
of occurring event.

It remains to choose the common constant $b > 0$ in (\ref{eq000010}). This may be determined by a non-statistical constraint
in censoring sensor networks that the average fraction of transmitting sensors at any
time step  is restricted to be at most $\eta \in (0,1)$  when no change occurs.
In this case, when no event occurs, the average fraction of transmitting sensors at any time step $n$ is
\begin{eqnarray*}
\frac{1}{K} \sum_{k=1}^{K} \Prob^{(\infty)}(U_{k,n} \ne \mbox{NULL})
= \frac{1}{K} \sum_{k=1}^{K} \Prob^{(\infty)}(W_{k,n} \ge \rho_{k} b
) \le \frac{1}{K} \sum_{k=1}^{K} \exp(-\rho_{k} b) \le \exp(-\rho_{\min} b),
\end{eqnarray*}
where $\rho_{\min} = \min_{1 \le k \le K} \rho_{k}$ and the second-to-last inequality follows from the well-known properties of the local CUSUM statistics that $\Prob^{(\infty)}(W_{k,n} \ge a) \le \exp(-a)$
for all $a >0,$ see, for example, Appendix 2 on Page 245 of Siegmund \cite{siegmund:1985}.
Thus a choice of $b = (1/\rho_{\min}) \log \eta^{-1}$ will guarantee that on average, at
most $100\eta\%$ of $K$  sensors will transmit messages
at any given time when no event occurs.

A special case occurs when all $K$ sensors are homogeneous in the sense that the KL information numbers  $I(g_{k}, f_{k})$'s in (\ref{eq000001}) are the same for all $k.$ Then we have $\rho_{\min} = 1/K,$ and our proposed choice of the local censoring parameter is given by
\begin{eqnarray} \label{eqnlocal}
b_{k} = \rho_{k} b= (1/K)(K \log \eta^{-1}) = \log \eta^{-1},
\end{eqnarray}
for all $k=1, \ldots, K.$ It is interesting to see that as the number $K$ of homogeneous sensors increases, the weight $\rho_{k}$ of each local sensor in the overall final decision is decreasing, but the common constant $b$ is increasing.
Thus the choice of the local censoring parameter $b_{k}$'s in (\ref{eqnlocal}) remains as a constant, and this seems   attractive to practitioners in censoring sensor networks.

It is important to emphasize for each of our proposed schemes $T(a)$ in (\ref{eq000012})-(\ref{rankCUSUM}),
the stopping time $T(a)$ is increasing as a function of the censoring parameters $b_{k}$'s when the global threshold value $a$ is {\it given}. That is, a larger value of $b_{k}$'s implies both larger ARL to false alarm and larger detection delays.
However, the situation becomes completely different when $T(a)$ is required to satisfy the false alarm constraint (\ref{eq000002}).
This is because different global threshold values $a$'s are needed for these schemes with different $b_{k}$'s,
and thus larger values of  $b_{k}$'s may or may not
lead to larger detection delays. Also see our numerical simulations below.

\subsection{Numerical Simulations}

In this subsection we report our numerical simulation results to illustrate the
usefulness of the proposed schemes  in (\ref{eq000012})-(\ref{rankCUSUM}). Suppose that there are $K= 100$ independent and
identical sensors in a system, and the observations at each sensor
are iid with mean $0$ and variance $1$ before the change and
with mean $1$ and variance $1$ after the change if affected. In our simulation study,
we simply assume that the change is  instantaneous if a sensor is affected, but we do not know which subset of sensors
will be affected by the occurring event.

For the purpose of comparison, we conduct numerical simulations for six families of global monitoring schemes:
\begin{itemize}
  \item  the ``MAX" scheme $T_{\max}(a)$ in (\ref{eq3n02}),
  \item  the ``SUM" scheme $T_{\rm sum}(a)$ in (\ref{eq3n03}),
  \item the order thresholding scheme $N_{order, r}(a)$ in (\ref{rankCUSUM}) with $r=10,$
  \item  the hard thresholding scheme $N_{hard}(a)$ in (\ref{eq000012}),
 \item  the soft thresholding scheme $N_{soft}(a)$ in (\ref{eq000012b}),
\item the combined thresholding schemes $N_{comb, r}(a)$ in (\ref{rankCUSUM1}) with $r=10.$
\end{itemize}

The first three schemes require all local sensors to send  all local CUSUM statistics $W_{k,n}$'s values to the fusion center at each and every time step, and corresponds to the case when  the local censoring parameter $b_{k} \equiv 0$ for all $k=1, \cdots, K.$ For order-thresholding in the families of $N_{order, r}(a)$ and $N_{comb, r}(a),$ we choose $r= 10$ to better understand the scenario when $10$ out of $100$ sensors are affected by the occurring event. For each of the last three schemes in the list, i.e., our three proposed schemes (\ref{eq000012})-(\ref{rankCUSUM1}), we further consider three different values of the local censoring parameters $b_{k}$'s:
\begin{description}
  \item[(i)] $b_{k} \equiv 1/2 \approx -\log(0.607)$ for all $k,$
  \item[(ii)] $b_{k}\equiv - \log(0.1) = 2.3026$  for all $k,$
  \item[(iii)] $b_{k}\equiv -\log(0.01)=4.6052$ for all $k.$
\end{description}
The choices of these values will guarantee
that when no event occurs, on average at most $\eta = 60.7\%, 10\%,$ and $1\%$
of $K=100$ homogeneous sensors will transmit messages at any given time, respectively.
Therefore, there are a total of $3+3*3 = 12$ specific schemes in our numerical simulation study.

\medskip
\begin{table}
 \centering

  \caption{A comparison of the detection delays of six families of schemes with $\gamma = 5000.$
  The smallest and largest standard errors of these $12$ schemes are also reported under each post-change hypothesis
  based on $2500$ repetitions in Monte Carlo simulations.
  }\label{table01}

  \bigskip

  \begin{tabular}{|l|c |c|c| c|c|c|c|c|c|}
    \hline
  &\multicolumn{9}{|c|}{\# sensors affected} \\
\cline{2-10}
         & 1& 3& 5  & $8$ & $10$ &$20$  & 30&$50$ & $100$ \\ \hline
 \hline
%   \multicolumn{9}{|c|}{Xie and Siegmund's schemes $T_{XS}(a,p_0)$ in (\ref{eqn_XS})}\\
%\hline
% $T_{XS}(a=53.5, p_0=1)$   &52.3& 18.7 & 12.2 &  & 6.7 &  & 2.3 & 2.0 \\
% $T_{XS}(a=19.5, p_0=0.1)$ &31.6& 14.2 & 10.4 &  & 6.7 &  & 2.8 & 2.0 \\
%\hline
%\hline  \multicolumn{9}{|c|}{Standard errors of all schemes below}\\ \hline
       Smallest standard error            & $0.18$ & $0.07$ & $0.05$ & $0.03$ & $0.03$ & $0.02$ & 0.01 &$0.01$ & $0.00$ \\ \hline
        Largest standard error            & $0.35$ & $0.12$ & $0.07$ & $0.06$ & $0.05$ & $0.04$ & 0.03 &$0.03$ & $0.03$  \\ \hline\hline
            \multicolumn{9}{|c|}{Schemes with $b_{k} \equiv 0$} \\ \hline
      $T_{\max}(a=11.27)$                 & $23.3$ & $16.3$ & $14.4$ & $13.0$ & $12.4$ & $10.9$ &10.2&$9.5$ & $8.7$ \\  \hline
 $T_{\rm sum}(a=88.66)$                   & $52.1$ & $21.8$ & $14.7$ & $10.3$ & $8.7$ & $5.2$ &3.9 & $2.9$ & $2.0$ \\ \hline
 $N_{order, r=10}(a=44.11)$               & $34.1$ & $15.5$ & $11.2$ & $8.5$  & $7.5$ & $5.5$ &4.8& $4.1$ & $3.4$\\ \hline \hline

    \multicolumn{10}{|c|}{Schemes $N_{hard}(a)$ in (\ref{eq000012}) with different positive $b_{k}$'s} \\ \hline
      $N_{hard}(a=85.60, b_k=0.50)$       & $52.9$ & $21.9$ & $14.9$ & $10.3$ & $8.7$ & $5.2$ &4.0& $2.9$ & $2.0$ \\ \hline
      $N_{hard}(a=52.21, b_k=2.3026)$     & $50.6$ & $20.7$ & $13.8$ & $9.6$  & $8.2$ & $5.2$ &4.2& $3.2$ & $2.4$ \\ \hline
      $N_{hard}(a=26.31, b_k=4.6052)$     & $39.8$ & $16.0$ & $11.5$ & $8.8$  & $7.9$ & $5.9$ &5.2& $4.4$ & $3.8$ \\ \hline \hline
                \multicolumn{10}{|c|}{Schemes $N_{soft}(a)$ in (\ref{eq000012b}) with different positive $b_{k}$'s} \\ \hline
      $N_{soft}(a=63.92, b_k=0.50)$       & $48.2$ & $20.2$ & $13.7$ & $9.7$  & $8.2$ & $5.1$ &4.0& $3.0$ & $2.0$ \\ \hline
      $N_{soft}(a=21.56, b_k=2.3026)$     & $33.9$ & $15.4$ & $11.2$ & $8.5$  & $7.5$ & $5.3$ & 4.5 &$3.7$ & $3.0$ \\ \hline
     $N_{soft}(a=8.29, b_k=4.6052)$      & $25.2$ & $13.8$ & $11.1$ & $9.2$  & $8.4$ & $6.7$ & 5.9&$5.2$ & $4.4$ \\ \hline \hline
            \multicolumn{10}{|c|}{Schemes $N_{comb, r}(a)$ in (\ref{rankCUSUM1}) with $r=10$ and different positive $b_{k}$'s} \\ \hline
$N_{comb, r}(a=44.11, b_k=0.50)$          & $34.1$ & $15.5$ & $11.2$ & $8.5$  & $7.5$ & $5.5$ &4.8& $4.1$ & $3.4$ \\  \hline
 $N_{comb, r}(a=43.88, b_k=2.3026)$       & $38.5$ & $16.8$ & $11.7$ & $8.6$  & $7.5$ & $5.5$ & 4.7&$4.0$ & $3.3$ \\ \hline
 $N_{comb, r}(a=26.31, b_k=4.6052)$       & $39.8$ & $16.0$ & $11.5$ & $8.8$  & $7.9$ & $5.9$ &5.2& $4.4$ & $3.8$ \\ \hline
\hline
  \end{tabular}
\end{table}

For each of these $12$ specific schemes $T(a),$  we first find the appropriate values of the global threshold $a$ to satisfy the false alarm constraint $\E^{(\infty)}(T(a)) \approx \gamma = 5000$ (within the range of sampling error). Next, using the
obtained global threshold value $a,$ we simulate the detection delay when the change-point occurs at time $\nu=1$ under several different post-change scenarios, i.e., different number of affected sensors. All Monte Carlo simulations are based on $m=2500$  repetitions.

%Given the fact that there are many possible post-change scenarios, we summarize our simulated results in two tables: Table 1 for the cases when $20$ or more sensors are affected and Table 2 for the cases when $10$ or less sensors are affected. In each table, two different values of the false alarm constraint $\gamma,$ $5*10^3$ and $5*10^4,$ are considered, and the detection delay is recorded as the Monte carlo estimate $\pm$ standard error. Also within the families of schemes $N_{hard}(a)$ and $N_{comb, r=10}(a)$, we report the numerical results of these specific schemes in order of increasing values of the censoring parameters $b_{k}$'s.

%\medskip
%{\bf\large ALL discussions below are from an earlier draft, and need to be rewritten.}

Table \ref{table01} summarizes our simulated detection delays of these $12$ schemes under $8$ different post-change hypothesis, depending on
 the number of affected sensors.   From Table \ref{table01}, among these $12$ specific schemes,
when a small number ($1 \sim 3$) of $100$ homogeneous sensors are affected by the event,
the  ``MAX" scheme $T_{\max}(a)$ is the best (in the sense of smallest detection delay),
the  ``SUM" scheme $T_{\rm sum}(a)$
is the worst, and all other schemes  are in-between. Similarly,
when a large number ($20$ or more) of $100$ homogeneous sensors  are affected,
the order is reserved: $T_{\rm sum}(a)$ is the best, $T_{\max}(a)$ is the worst, and
all other schemes are in-between. However, when $5 \sim 10$ sensors are affected, the schemes with
order-thresholding $r=10$  yield the smallest detection delays, since they are designed
to detect the scenario when $10$ sensors are affected by the event. An interesting observation
is that the soft-thresholding scheme $N_{soft}(a)$ can also yield the smallest
detection delays with a suitable choice of $b_{k}$'s. In addition,
it is clear from  Table \ref{table01} that for each given scheme, the fewer affected sensors we have,
the larger detection delay it will have. All these results are consistent with our intuition.

It is worth emphasizing that for the families of the hard-thresholding schemes $N_{hard}(a)$ in (\ref{eq000012})
or the soft-thresholding schemes $N_{soft}(a)$ in (\ref{eq000012b}), a larger censoring value of $b_{k}$ actually leads
to a smaller detection delay when only a few sensors (between
$1$ and $5$ sensors) are affected. This suggests that a larger censoring value $b_{k}$ may actually be necessary for efficient detection when the
affected sensors are sparse.

A surprising and possibly counter-intuitive result in Table \ref{table01} is the effect of  not so large values of censoring parameters $b_{k}$'s in finite sample simulations. For instance,
the performances of the ``SUM" scheme $T_{\rm sum}(a)$
and the hard thresholding scheme $N_{hard}(a,b_{k}=0.50)$ are similar in view of sampling errors.
Likewise,  the top-$r$ thresholding  scheme $N_{order, r=10}(a)$ and the combined thresholding
scheme $N_{comb, r=10}(a,b_{k}=0.50)$ also have identical performances.
The interpretation in the censoring sensor networks context is as follows:
using our proposed communication policy in (\ref{hardthreshold}), we only need $\exp(- b_{k}) = \exp(-0.5)=60.7\%$ of $100$ sensors to transmit information to the fusion center at any given time when no event occurs, but we can still be as effective as the full transmission scenario when all sensors transmit information at all time steps. In other words, much communication costs can be saved by our proposed schemes $N_{hard}(a)$ or $N_{comb, r}(a)$ with  not so large  values of $b_{k}$'s.

It is also interesting to see the effect of the order-thresholding parameter $r$ in finite sample simulations when the hard-thresholding parameters $b_{k}$'s are large. From Table  \ref{table01}, when the false alarm constraint $\gamma$ in (\ref{eq000002}) is only moderately large, e.g., $\gamma = 5000,$ the performances of $N_{hard}(a, b_{k})$ and $N_{comb, r=10}(a,b_{k})$  are identical when $b_{k}= 4.6052$ --- they not only have the same global threshold $a,$ but also have the same detection delays. Intuitively, the stopping time $N_{comb,r}(a,b_{k})$ is decreasing as a function of $r,$ and thus we have $N_{hard}(a,b_{k}) = N_{comb, r=K}(a,b_{k}) \le N_{comb, r=10}(a,b_{k})$ when $b_{k}=4.6052.$ So one may wonder why our numerical simulations lead to identical results?
One explanation is that with such a choice of $b_{k}=4.6052,$ when no event occurs, on average there is at most $1$ non-zero sensor message  received in the fusion center at any given time, and thus there is little difference whether one uses the sum of the largest $r=10$ sensor messages or

uses the sum of all $K=100$ sensor messages. Hence similar performances are observed in finite-sample simulations.

%In summary, from the performance viewpoint, using one of hard-thresholding and top-$r$ thresholding approaches may be sufficient in certain applications, since the performance of the combined censoring scheme $N_{comb, r}(a,b)$ can be similar to that of either the hard-thresholding scheme $N_{hard}(a,b)$ or the top $r$-thresholding scheme $N_{order, r}(a),$ especially when the false alarm constraint $\gamma$ in (\ref{eq000002}) is only moderately large.

\subsection{Asymptotic Optimality Theory}

In this subsection, we provide theoretical justification of our choices of the local censoring parameters $b_{k}$'s in (\ref{eq000010})-(\ref{eq000011}) and we will show the corresponding schemes hold certain asymptotic optimality properties.  To emphasize the choices of  $b_{k} = \rho_{k} b$ in (\ref{eq000010})-(\ref{eq000011}) with $b \ge 0$ being the common constant, we rewrite our proposed schemes as $N_{hard}(a, b), N_{soft}(a,b)$ and $N_{comb,r}(a, b)$ in this subsection and only in this subsection.

%Note that the threshold $a$ in the soft-thresholding $N_{soft}(a)$ in (\ref{eq000012b}) has a different meaning as compared to those of the other three schemes.
%It is not easy to present a general asymptotic theory for all four proposed schemes in (\ref{eq000012})- (\ref{rankCUSUM}), since the threshold $a$ in the soft-thresholding $N_{soft}(a)$ in (\ref{eq000012b}) has a different meaning as compared to those of the other three schemes. To simplify our presentation, we will focus on the schemes $N_{hard}(a)$ in (\ref{eq000012}) and $N_{comb,r}(a)$ in (\ref{rankCUSUM1}), and the asymptotic properties of $N_{soft}(a)$ in (\ref{eq000012b}) will be similar to those of $N_{hard}(a^*)$ in (\ref{eq000012}) with a new threshold $a^*= a + \sum_{k=1}^{K} b_{k}.$ In addition,

Let us begin with a rigorous definition of the post-change hypothesis. We assume that the $k$-th data stream is affected at time $\nu_{k} = \nu + \delta_{k},$ where the term $\delta_{k} \in [0, \infty]$
denotes the delay of the occurring event's impact on the $k$-th data stream, and $\delta_{k}
= \infty$ implies that the $k$-th data stream
is not affected. That is, the density
function of the sensor observations $X_{k,n}$'s of the $k$-th data stream changes from $f_{k}$ to $g_{k}$ at time $\nu_{k} = \nu + \delta_{k}.$
In the case when the change is instantaneous, the delay effect $\delta_{k}$ only takes two possible values, $0$ or $\infty.$ Here we relax such an assumption a little bit, and assume that the change might not be instantaneous.

To simplify our arguments and highlight our main ideas, we will assume that the delay effects $\delta_{k}$'s satisfy the following post-change hypothesis set $\Delta:$
\begin{eqnarray} \label{eqset}
\Delta = \big\{(\delta_{1}, \ldots, \delta_{K}):  \mbox{the $\delta_{k}$'s either $=\infty$ or satisfy $0 \le \delta_{k} << \log \gamma$ } \mbox{and } \min_{1 \le k \le K} \delta_{k} =0 \big\}.
\end{eqnarray}
where $\gamma$ is the false alarm constraint in (\ref{eq000002}), and $x(t) << y(t)$ implies that $x(t)/y(t) \rightarrow 0$ as $t \rightarrow \infty.$ Note that the assumption of
$\min_{1 \le k \le K} \delta_{k} =0$ is trivial, since otherwise the system is actually affected by the occurring event at the ``new" change-point $\nu' = \nu + \min_{1 \le k \le K} \delta_{k}.$ The assumption of $\delta_{k} << \log \gamma$ is a technical assumption to ensure that one is able to utilize all affected data streams to raise a global alarm  subject to the false alarm constraint $\gamma$ in (\ref{eq000002}). In other words, we only consider the scenario when the differences on the finite delay affects $\delta_{k}$'s are not too large as compared to the typical order ($\log \gamma$)
of detection delays. A sufficient condition to satisfy this assumption is when all finite $\delta_{k}$'s are uniformly bounded by some constants that do not depend on  the false alarm constraint $\gamma$  in (\ref{eq000002}).

Next, let us define the detection delay of a global monitoring scheme rigorously when the event occurs at the unknown time $\nu$ with specific delay effects $\delta_{k}$'s.  Suppose a global monitoring scheme raises an alarm at time $T \ge \nu,$ it takes $T -\nu+1$ time steps from the post-change scenario to indicate that an event might occur, and thus $T - \nu + 1$ can be regarded as the detection delay. To take into account of the randomness of $T$ and the uncertainty of $\nu,$  a widely used rigourous  definition of the detection delay of $T$ is the following ``worst case" detection delay defined in Lorden
\cite{lorden:1971},
\begin{eqnarray*}
&&\overline{\E}_{\delta_{1}, \cdots, \delta_{K}}(T) = \quad \sup_{\nu \ge 1}\
\mbox{ess}\sup \E^{(\nu)} \Big( (T - \nu
+ 1)^{+} \Big| \mathcal{F}_{\nu-1} \Big).\nonumber
\end{eqnarray*}
Here the $\delta_{k}$'s are the delay effects,  $\mathcal{F}_{\nu-1} = (X_{1,[1,\nu-1]}, \ldots, X_{K, [1,
\nu-1]})$ denotes past global information at time $\nu,$
$X_{k,[1,\nu-1]}= (X_{k,1}, \ldots, X_{k,\nu-1})$ is past local
information for the $k$-th data stream, and
$\Prob^{(\nu)}$ and $\E^{(\nu)}$ denote
the probability measure and expectation when the event occurs at time $\nu.$

Mathematically, the problem of finding an efficient global monitoring scheme  can then be formally formulated as
finding a stopping time $T$ such that the detection delay $\overline{\E}_{\delta_1,  \ldots, \delta_{K}}(T)$
is as small as possible for all possible combinations of $(\delta_1, \cdots, \delta_{K}) \in \Delta$ in (\ref{eqset})
subject to the false alarm constraint (\ref{eq000002}).

We are now ready to present the asymptotic optimality properties of our proposed schemes, $N_{hard}(a, b),$ $N_{soft}(a, b),$ $N_{order, r}(a),$ and $N_{comb,r}(a, b),$
under the standard asymptotic setting in which the number of data streams $K$ is fix and the false alarm constraint
$\gamma$ goes to $\infty.$ Later we will briefly add some general remarks, including the properties when
both the number $K$ of data streams and the false alarm constraint $\gamma$ go to $\infty$ in some appropriate rates.

The following theorem, whose proof is postponed to Section 5, derives the information bound on the detection delays of any globally monitoring  schemes when $\Delta$ is defined in (\ref{eqset}), as the false alarm constraint $\gamma$ in (\ref{eq000002}) goes to $\infty.$
\begin{theorem} \label{theorem1}
Assume a scheme $T(\gamma)$ satisfies the false alarm constraint (\ref{eq000002}). Then for any given post-change hypothesis
$(\delta_{1},\ldots, \delta_{K}) \in \Delta,$ as $\gamma$ goes to $\infty,$
\begin{eqnarray} \label{eqlowerbound}
\overline{\E}_{\delta_1,  \ldots, \delta_{K}}(T(\gamma)) \ge (1+o(1)) \frac{\log\gamma}{J(\delta_1, \ldots, \delta_{K})},
\end{eqnarray}
where
\begin{eqnarray} \label{Jinfo}
J(\delta_1, \ldots, \delta_{K}) = \sum_{k=1}^{K} I(g_{k}, f_{k}) I\{\delta_{k} < \infty\},
\end{eqnarray}
and $I(g_{k}, f_{k})$ is the KL information number defined in (\ref{eq000001}), and $I\{A\}$ is the indicator function of set $A.$
\end{theorem}

\bigskip
Next, when the local detection statistics $W_{k,n}$'s are the local CUSUM statistics in (\ref{eq3n01}) and the local censoring parameters are defined by $b_{k}=\rho_{k} b$'s in (\ref{eq000010})-(\ref{eq000011}) for some common constant $b \ge 0,$  we establish the asymptotic properties of
 our proposed schemes, $N_{hard}(a, b)$ in (\ref{eq000012}), $N_{order, r}(a)$ in (\ref{rankCUSUM}), and $N_{comb,r}(a, b)$ in (\ref{rankCUSUM1}),
as the global threshold $a$ goes to $\infty,$ regardless of the false alarm constraint (\ref{eq000002}).  The proof of the following theorem  is presented in detail in Section 5.
\begin{theorem} \label{theorem2}
As $a \rightarrow \infty,$ let $b'=b'(a)$ be a constant such that both $b'$ and $a - b'$ go to $\infty.$
\begin{description}
  \item[(i)]   The hard-thresholding scheme $N_{hard}(a,b)$  in (\ref{eq000012}) satisfies
\begin{eqnarray} \label{eq000006}
\E^{(\infty)}(N_{hard}(a,b)) \ge \frac{e^{a}}{1+a+\frac{a^2}{2!}+ \cdots+ \frac{a^{K-1}}{(K-1)!}}.
\end{eqnarray}
for any real number $b \ge 0.$   Moreover, for any combination $(\delta_1, \ldots, \delta_{K}) \in \Delta$  defined in (\ref{eqset}), and for all $0 \le b \le b',$ we have %as both $b$ and $a - b$ go to $\infty,$
\begin{eqnarray} \label{eq0011}
& & \overline{\E}_{\delta_1, \ldots, \delta_{K}}(N_{hard}(a,b)) \le
\frac{a}{ J(\delta_1, \ldots, \delta_{K}) }+ O(\sqrt{b}) + O(1) + O\Big(\max_{\delta_{k}: \delta_{k} < \infty} (\delta_{k})\Big),
\end{eqnarray}
where the maximum is taken over all those $\delta_{k}$'s that are finite, and $J(\delta_1, \ldots, \delta_{K})$ is defined
in (\ref{Jinfo}).
 \item[(ii)]   The soft-thresholding scheme $N_{soft}(a,b)$  in (\ref{eq000012b}) satisfies relation (\ref{eq000006}) for all $b \ge 0.$
 Moreover, it also satisfies relation (\ref{eq0011}) except that the $a$ in the right-hand side of (\ref{eq0011}) is replaced by $a + b.$
  \item[(iii)] For any integer $1 \le r \le K,$ the order-$r$ thresholding scheme $N_{order, r}(a)$ in (\ref{rankCUSUM}) and the combined thresholding scheme $N_{comb, r}(a,b)$ in (\ref{rankCUSUM1}) with $b \ge 0$ also satisfy relation (\ref{eq000006}). In addition, for $0 \le b \le b',$  both schemes satisfy (\ref{eq0011}) whenever $\sum_{k=1}^{K} I\{\delta_{k} < \infty\} \le r,$ i.e., when the occurring event affects at most $r$ sensors.
\end{description}
\end{theorem}

\bigskip
Finally, when the local detection statistics $W_{k,n}$'s are the local CUSUM statistics in (\ref{eq3n01}) and the local censoring parameters $b_{k}$'s are defined in (\ref{eq000010})-(\ref{eq000011}) with $b \ge 0$ being the common constant, the asymptotic optimality properties of our proposed schemes can be summarized as follow.

\begin{corollary} \label{corollary001}
For a given $K$ and for any $b \ge 0,$ with the choice of
\begin{eqnarray} \label{eq000008}
a= a_{\gamma} = \log \gamma + (K-1+o(1)) \log\log \gamma,
\end{eqnarray}
the hard-thresholding scheme $N_{hard}(a_{\gamma},b)$ satisfies the false alarm
constraint (\ref{eq000002}).  Moreover, if $b' = b_{\gamma}'$ is chosen such
that both $b'$ and $a- b'$ go to
$\infty$ as $\gamma$ go to $\infty,$ then for all $0 \le b \le b',$
\begin{eqnarray*}
\overline{\E}_{\delta_1, \ldots, \delta_{K}}(N_{hard}(a,b)) \le \frac{ \log\gamma + (K-1
+ o(1)) \log\log\gamma}{  J(\delta_1, \ldots, \delta_{K}) } +
O(\sqrt{b}) + O(1)
\end{eqnarray*}
for all possible post-change hypothesis $(\delta_1, \ldots, \delta_{K}) \in \Delta$ in (\ref{eqset}). Therefore, for any given $b \ge 0,$ the hard-thresholding schemes $N_{hard}(a,b)$ in (\ref{eq000012})
asymptotically minimize $\overline{\E}_{\delta_1, \ldots, \delta_{K}}(N_{hard}(a,b))$ (up to
the first-order) for each and every post-change hypothesis
$(\delta_1, \ldots, \delta_{K}) \in \Delta$ subject to the false alarm
constraint (\ref{eq000002}), as  $\gamma$ in (\ref{eq000002}) goes to $\infty.$
The conclusion also holds if $N_{hard}(a,b)$ is replaced by either the order-thresholding scheme $N_{order,r}$ in (\ref{rankCUSUM}) or
the combined thresholding scheme $N_{comb, r}(a,b)$ in (\ref{rankCUSUM1}) when the occurring event affects at most $r$ data streams, i.e.,
when $(\delta_1, \ldots, \delta_{K}) \in \Delta$ satisfies $\sum_{k=1}^{K} I\{\delta_{k} < \infty\} \le r.$
\end{corollary}

\medskip
{\bf Proof:} This corollary follows at once from Theorems \ref{theorem1} and
\ref{theorem2}. In particular, the choice of $a_{\gamma}$ in (\ref{eq000008})
follows from (\ref{eq000006}) and the fact that $1+a+\frac{a^2}{2!}+ \cdots+ \frac{a^{K-1}}{(K-1)!}
\sim \frac{a^{K-1}}{(K-1)!}$ if $K$ is fixed and $a$ goes to $\infty.$ \hspace*{\fill}~\QED

\bigskip
It is worth pointing out several implications of our asymptotic
results. First of all, from Corollary \ref{corollary001}, it is
interesting to note that the first-order term of the detection delays of the hard-thresholding scheme $N_{hard}(a_{\gamma},b_{\gamma})$
is $(\log\gamma)/ J(\delta_1, \ldots, \delta_{K}),$ the asymptotic lower bound in (\ref{eqlowerbound}) in Theorem \ref{theorem1}, but
its second-order term contains both $O(\log\log\gamma)$ and $O(\sqrt{b}).$
Hence, as the common constant $b$ changes from $0$ to $b_{\gamma}'=\log \gamma (\sim a_{\gamma}),$  the second-order term
of the detection delays changes from $O(\log\log \gamma)$ to $O(\sqrt{\log \gamma}).$ Hence, if we want to keep the second-order term of the
detection delay to be as small as the order of $O(\log\log \gamma)$ for each and every
possible post-change hypothesis (i.e., different combination of
affected data streams), then the maximum choice of $b$ should be
$b= O((\log\log\gamma)^2) = O( (\log a_{\gamma})^2).$

Second,  recall that relations (\ref{eqnheuris0}) and (\ref{eqnheuris}) provide heuristic choices of the global threshold $a$ based on Chebyshev's inequality and the CLT approximation, respectively. For the purpose of better understanding these heuristic choices, below we will apply the spirit of these approximations to relation (\ref{eq000006}) when $K$ is large. Note that the right-hand side of (\ref{eq000006}) is just $1/\Prob(U_{K} \ge a),$ where $U_{K}$ denotes the sum of $K$ iid exponential random variables with mean $1.$ To estimate the small value $\Prob(U_{K} \ge a)$ for large $K,$ one way is to use the CLT that leads to $(U_{K} - K)/\sqrt{K} \sim N(0,1).$ A choice of $a_{\gamma} \approx K + z_{1/\gamma}  \sqrt{K}$ will yield $\Prob(U_{K} \ge a) = 1/\gamma,$ and thus the right-hand side of (\ref{eq000006})
satisfies the global false alarm constraint in (\ref{eq000002}). This is consistent with the heuristic choice of $a$ in (\ref{eqnheuris}).

The other way is to use Chebyshev's inequality and the theory of large deviations: for
any constant $w =a /K > 1,$ we have
\[
\lim_{K \rightarrow \infty} -\frac{1}{K} \log \Prob(U_{K} \ge a) = \lim_{K \rightarrow \infty} -\frac{1}{K} \log \Prob( \frac{1}{K} U_{K} \ge w)= w-1-\log(w),
\]
see, for example, Durrett \cite[Ch. 1.9]{durrett:1996}. Hence,  when the global false alarm constraint $\gamma$  in (\ref{eq000002}) and the dimension $K$ go to $\infty$ simultaneously in such a way that $(\log \gamma) / K = w-1-\log(w)$ is constant,
a choice of the threshold $a = w K = (\log\gamma)w/(w-1-\log w)$ will lead the right-hand side of (\ref{eq000006})
satisfy the global false alarm constraint in (\ref{eq000002}). This is similar to the choice of $a$ in (\ref{eqnheuris0}).

We should mention that given the above heuristic choices of $a,$ a comparison of relation (\ref{eq0011}) in Theorem \ref{theorem2} with  the lower bound in Theorem \ref{theorem1} for fixed $K$ and large $\gamma$ suggests that our proposed schemes may no longer achieve the lower bound in Theorem \ref{theorem1}, which may or may not provide a sharp lower bound on  the detection delays as both the dimension $K$ and the global false alarm constraint $\gamma$ go to $\infty$ simultaneously in a suitable rate.

%, the asymptotic efficiency of our proposed schemes as compared to the lower bound in Theorem \ref{theorem1} will be $w/(w-1-\log w),$ which goes to $1$ if $w$ goes to $\infty,$ or equivalently, if $w - 1 - \log(w) = (\log \gamma)/ K$ goes to $\infty.$ In other words, the asymptotic optimality properties of our proposed  schemes seem to still hold when  $(\log \gamma) / K$ goes to $\infty.$ However, in other scenarios, our proposed schemes no longer achieve the asymptotic lower bound in Theorem \ref{theorem1}, which may or may not provide a sharp lower bound on  the detection delays.

Third, let us further elaborate the communication rate between sensors and the fusion center in the context of monitoring $K$ homogeneous sensors.
As mentioned in (\ref{eqnlocal}), if we want at most $100\eta\%$ of $K$ homogeneous sensors on average to transmit messages to the fusion center
at any given time when no event occurs, we can choose the local censoring parameter $b_{k} = \log \eta^{-1},$ and
thus the common censoring constant of our proposed schemes will be $b=K \log \eta^{-1}$ for
a given $\eta \in (0,1).$  Meanwhile, in our theorems and corollary, the asymptotic optimality properties of our proposed schemes
hold under the condition that $(a - b)$ goes to $\infty.$ When the global threshold $a$  satisfies
$a = a_{\gamma} \approx K + \sqrt{K} z_{1/\gamma}$ as stated in the previous remark, then the condition of $a - b \rightarrow \infty$
is equivalent to  $\log \eta^{-1} \le 1,$ i.e.,  $\eta \ge 1/e = 36.8\%.$  In other words, when
at least $36.8\%$ of $K$ homogeneous sensors can transmit messages at any given time when no
event occurs, we can still develop efficient global monitoring schemes (e.g.,
$\{N_{hard}(a,b)\}$) that are asymptotically optimal to
detect each and every possible combination of affected data streams.
However, if $\eta < 36.8\%,$ then it is unclear whether our proposed schemes  can still
effectively detect all different possible post-change hypotheses.
Also see our numerical simulations in the previous section.

%Last but not the least, our theorems are proved when local CUSUM statistics are used, and it is natural to ask whether the results also hold if local Shiryaev-Roberts statistics  (in the logarithm scale) are used, since it is well-known that the CUSUM and Shiryev-Roberts schemes have similar performance when monitoring one-dimensional data stream. Our numerical simulation study seems to suggest that the answer is a ``Yes," as the Shiryev-Roberts like version also has similar performance as the CUSUM version when monitoring a large number of data streams. Unfortunately, we are unable to prove rigorously that the Shiryev-Roberts like version of our proposed schemes is asymptotically optimal, and the major challenge is to establish a lower bound on the average run length to false alarm in parallel to those in (\ref{eq000006}).

\section{A Second Example: Normal Distribution with Unknown Post-Change Means}

Suppose that we are monitoring $K$ data streams $X_{k,n}$'s in (\ref{eqn01}). Initially, the data $X_{k,n}$'s are iid $N(0,1).$ At some unknown time $\nu,$ an occurring event may change the distribution of the $k$-th local data stream  to $N(\mu_{k},1)$ for some unknown $1 \le k \le K$. As in the previous section, we do not know which subset of local data streams are affected, but here we add a new challenge that we do not know the values of the post-change means $\mu_{k}$'s when affected.  We want to develop a system-wise online monitoring scheme that can detect the change as soon as possible, subject to the global false alarm constraint $\gamma$ in (\ref{eq000002}).

Xie and Siegmund \cite{xie2013sequential} investigates this problem under the assumption that the post-change mean $\mu_{k} > 0$ for all $k.$ By assuming that the fraction  $p_0$ of affected data stream is known, the main scheme they proposed is motivated from a semi-Bayesian approach and is defined by
\begin{eqnarray} \label{eqn_XS}
T_{XS}(a, p_0) =
\inf\left\{n \ge 1: \max_{0\le i < n} \sum_{k=1}^{K} \log(1-p_0+ p_0 \exp\Big[ \big(U_{k,n,i}^{+}\big)^2/2\Big] \ge a \right\}.
\end{eqnarray}
where  for all $1 \le k \le K, 0 \le i < n,$
\[
U_{k,n,i}^{+}= \max\big(0, \frac{1}{\sqrt{n-i}} \sum_{j=i+1}^n X_{k,j}\big)
\]
%As one can see, this global monitoring scheme requires one to store all of past observations in memory at each time step $n,$ and cannot be implemented recursively.
Some simplified versions have also been proposed to reduce the memory requirement to a large window of the most recent observations. However, all schemes in Xie and Siegmund \cite{xie2013sequential} are not suitable in the context of censoring sensor networks in Figure \ref{fig:mei1}: besides being computationally expensive, the implementation of their schemes requires the fusion center to have full access to all data streams at each time step.

%computationally expensive since they cannot be implemented recursively over time, and thus they are not suitable for  monitoring large-scale data streams over long period time $n.$ In addition, as mentioned in Xie and Siegmund \cite{xie2013sequential}, it is very important to consider the positive part $U_{k,n,i}^{+}$ as this plays the role of dimension reduction when the post-change means are known to be positive. Thus their methodologies are not designed in the scenario when the post-change means can be positive or negative.

It has been an open problem to develop a scalable global monitoring scheme  in  the censoring sensor networks context that can detect both positive and negative local mean shifts for affected local data streams. Part of the reason is that for the $K$ local data streams, there are $2^{K}$ potential different combinations of positive or negative local shifts, which is huge for a large $K.$
% Indeed, no scalable schemes have been proposed for monitoring a large $K$ number of data streams when the local shifts can be positive or negative.

In this section, we  illustrate how to tackle this open problem based upon our proposed SUM-shrinkage statistics in (\ref{shrinkage}).  The  main challenge is to choose a suitable local detection statistic $W_{k,n}$ that can be easily computed and has the ability to detect both positive and negative local mean shifts. Once such local detection statistic $W_{k,n}$'s are defined, it is evident from the previous section that we can use any shrinkage transformation such as hard-thresholding, soft-thresholding, or order-thresholding to develop a global monitoring scheme. Below we use the soft-thresholding transformation as a demonstration. Our numerical simulation experience suggests that as a continuous function, the soft-thresholding transformation often yields smaller detection delays than the hard-thresholding transformation, and is computationally more efficient than the order-thresholding transformation.

To be more specific, in this section we will consider the soft-thresholding scheme
\begin{eqnarray} \label{eqn:soft}
N_{soft}(a)
%= \inf\left\{n:\sum_{k=1}^K G_n\geq d\right\} \nonumber \\
=\inf\left\{n \ge 1:\sum_{k=1}^{K} \max(W_{k,n}- b_1, 0) \geq a \right\},
\end{eqnarray}
where,  for simplicity, all  transformations $h_{k}$'s are chosen to the same soft-thresholding transformation $\max(u-b_1, 0)$ for some constant $b_1 > 0.$ Our focus is  how we can construct the local detection statistics $W_{k,n}$'s suitably.

The remainder of this section is as follows. Subsection 4.1 reviews  the recursive register approach of Lorden and Pollak \cite{lorden2008sequential} for monitoring a single data stream, which is adapted to monitoring positive and negative mean shifts in Subsection 4.2. Subsection 4.3 provides a Bayesian interpretation of the soft-thresholding scheme  as well as an efficient numerical algorithm of our proposed SUM-shrinkage scheme that only uses fixed $6 K$ registers to store all past information and involves $O(K)$ computations at each given time step $n.$ Numerical simulation results are summarized in subsection 4.4.

\subsection{The Recursive Register Approach  of Lorden and Pollak \cite{lorden2008sequential}} \label{sec:method0}

To abuse notation, in this subsection we suppress the subscript $k$ of the $k$-th data stream, and consider the local monitoring problem with respect to the one-dimensional data stream $\{X_1, X_2, \ldots\}$ whose distribution may change from $N(0,1)$ to $N(\mu, 1)$ with unknown post-change mean $\mu$ at some unknown time $\nu.$  Lorden and Pollak \cite{lorden2008sequential} focuses on the case when the unknown post-change mean $\mu > 0,$ and makes a technical assumption that $\mu \ge \rho,$ where $\rho \ge 0$ is the smallest mean shift that is meaningful in practice, e.g. $\rho=0.25.$

A high-level description of the recursive register approach  of Lorden and Pollak \cite{lorden2008sequential} is as follows. Recall that the CUSUM statistics are defined in (\ref{eq3n01}), and for one-dimensional normal distributed data, the CUSUM statistics have a simpler recursive form
\begin{eqnarray} \label{eqnCUSUM}
W_{n} = \max(W_{n-1} + \mu X_{n} - \frac12 \mu^2, 0)
\end{eqnarray}
and $W_0 = 0.$ When $\mu$ is unknown, we can continue to use this recursive formula to define a detection statistic if we replace the true unknown $\mu$ by its estimate from the past observed data. A key observation in Lorden and Pollak \cite{lorden2008sequential} is that at each given time step $n,$ the CUSUM-type detection statistics can produce a candidate post-change time $\hat \nu \in \{0, 1, \cdots, n-1\},$ and thus the observations $X_{\hat \nu}, X_{\hat \nu + 1}, \cdots, X_{n-1}$ can be used to estimated the post-change mean $\mu$ in (\ref{eqnCUSUM}). Specifically, at any given time step $n,$
define $\hat \nu$ as the largest $0 \le i \le n-1$ such that $W_{i} = 0,$ and denote by $T_n$ and $S_n$  the total number and the summation of observations $X_{i}$'s  between the candidate post-change time $\hat \nu$ and time step $n-1.$ That is,
\begin{equation} \label{eqn:TS}
T_{n} = n - \hat \nu \qquad \mbox{ and } \qquad S_{n} = \sum_{i=\hat \nu}^{n-1} X_{i}.
\end{equation}
By the method of moments estimator or maximum likelihood estimator method, the post-change mean $\mu$ can be estimated by $S_{n} /T_{n}$ at time step $n.$ If we treat the pre-specified nonnegative constants $s$ and $t$ as a prior, then a Bayes-type estimate of $\mu$ is $\hat \mu_{n} = (s+S_{n}) / (t+T_{n}),$ which includes $S_{n} / T_{n}$ as a special case when $s=t=0.$ After taking into account that $\rho$ is the smallest post-change mean we are interested in, one can estimate $\mu$ at time step $n$ by
\begin{eqnarray} \label{eqn:muhat}
\hat \mu_{n} = \max\Big(\rho, \frac{s+S_{n}}{t+T_{n}} \Big).
\end{eqnarray}

From the algorithm viewpoint, the recursive register approach of Lorden and Pollak \cite{lorden2008sequential} can be recursively implemented as follows. Let $S_0 = T_0 = W_0 =X_0 = 0,$ and $\hat \mu_{1} = \rho.$ For all $n \ge 1,$
\begin{equation} \label{eq:W_n}
W_n = \max\left(W_{n-1}+ \hat\mu_{n} X_{n} - \frac12 (\hat \mu_{n})^2, 0\right),
\end{equation}
where $\hat \mu_{n}$ is defined in (\ref{eqn:muhat}) and
\begin{eqnarray} \label{eq:STn}
\left(
\begin{array} {rr}
S_n\\T_n
\end{array}
\right)
=
\left\{
\begin{array}{ll}
\left(\begin{array}{ll}S_{n-1}+X_{n-1} \\ T_{n-1}+1 \end{array}\right) &\mbox{if $W_{n-1}>0$}\\
\left(\begin{array}{rr}0 \\ 0 \end{array}\right) &\mbox{if $W_{n-1}=0$}.\end{array}
\right.
\end{eqnarray}
In other words, the local detection statistics $W_{n}$'s can be computed recursively as the part of three-dimensional vectors $(S_n, T_n, W_n),$ or  four-dimensional vectors $(S_n, T_n,  \hat \mu_{n}, W_n).$ It is important to note that $(S_{n}, T_{n}, \hat \mu_{n})$ only uses the observations up to time $n-1$ for the purpose of estimating the post-change mean $\mu,$ so that the data $X_{n}$ is reserved for the local detection statistics $W_{n}$ for the purpose of detecting changes. It was shown that the detection scheme based on the detection statistic $W_{n}$ in (\ref{eq:W_n}) is asymptotically optimal whenever the true post-change mean $\mu \ge \rho > 0,$  see Theorems 3.1-3.3 of Lorden and Pollak \cite{lorden2008sequential}.

%\zhang{Should we add something about the scenario when we need to estimate negative post-change mean $\mu < -\rho$ and the estimator
%\begin{eqnarray} \label{eqn:nemuhat}
%\hat \mu_{n} = \max\Big(-\rho, \frac{-s+S_{n}}{t+T_{n}} \Big).
%\end{eqnarray}
% }

\subsection{Our Proposed Local Detection Statistics $W_{k,n}$'s} \label{sec:method}

%We now illustrate how to extend the scheme of Lorden and Pollak \cite{lorden2008sequential} to our context of monitoring large-scale data streams.

Since we are interested in detecting both positive and negative local mean shifts for affected data streams, we propose to extend the detection statistic $W_{n}$ in (\ref{eq:W_n}) of Lorden and Pollak \cite{lorden2008sequential} from one-sided to two-sided.
Observe that detecting negative local mean shift of $X_{k,n}$'s is equivalent to detecting positive local mean shift of  $-X_{k,n}$'s, we propose the following two-sided local detection statistic for each local data stream at time $n:$
\begin{eqnarray} \label{eq:W_n2}
W_{k,n}=\max⁡(W_{k,n}^{(1)}, W_{k,n}^{(2)}),
\end{eqnarray}
where $W_{k,n}^{(1)}$ and $W_{k,n}^{(2)}$ are the local detection statistics of Lorden and Pollak \cite{lorden2008sequential} for detecting positive and negative mean shifts, respectively. Specifically,
\begin{eqnarray} \label{eq:W_n3}
W_{k,n}^{(1)} = \max\left( W_{k,n-1}^{(1)} + \hat\mu_{k,n}^{(1)} X_{k,n} - \frac12 (\hat \mu_{k,n}^{(1)})^2, 0\right), \cr
W_{k,n}^{(2)} = \max\left( W_{k,n-1}^{(2)} + \hat\mu_{k,n}^{(2)} X_{k,n} - \frac12 (\hat \mu_{k,n}^{(2)})^2, 0\right),
\end{eqnarray}
where
\begin{eqnarray} \label{eqn:muhat2}
\hat \mu_{k,n}^{(1)} = \max\Big(\rho, \frac{s+S_{k,n}^{(1)}}{t+T_{k,n}^{(1)}} \Big) > 0, \qquad
\hat \mu_{k,n}^{(2)} = \min\Big( - \rho, \frac{-s+S_{k,n}^{(2)} }{t+T_{k,n}^{(2)}} \Big) < 0,
\end{eqnarray}
and for $j=1,2$ and for any $k,$ the sequences  $(S_{k,n}^{(j)}, T_{k,n}^{(j)})$ are defined recursively
\begin{eqnarray} \label{eq:STn2}
\left(
\begin{array} {rr}
S_{k,n}^{(j)}\\T_{k,n}^{(j)}
\end{array}
\right)
=
\left\{
\begin{array}{ll}
\left(\begin{array}{ll}S_{k,n-1}^{(j)}+X_{k, n-1} \\ T_{k,n-1}^{(j)}+1 \end{array}\right) &\mbox{if $W_{k,n-1}^{(j)} >0$}\\
\left(\begin{array}{rr}0 \\ 0 \end{array}\right) &\mbox{if $W_{k,n-1}^{(j)}=0$}\end{array}
\right.
\end{eqnarray}

Note that $\hat \mu_{k, n}^{(1)}$ and $\hat \mu_{k, n}^{(2)}$ in (\ref{eqn:muhat2}) are the estimates of the post-change mean when restricted to the positive and negative values, respectively, under the assumption that $|\mu| \ge \rho.$ Clearly, $W_{k,n}^{(1)}$ is designed to detect positive local mean shift, whereas  $W_{k,n}^{(2)}$ is to detect negative local mean shifts. Also the two-sided local detection statistic $W_{k,n}$ in (\ref{eq:W_n2}) is always nonnegative for any $k$ at any time step $n$, and it will become large when there is a local mean shift no matter whether such mean shift is positive or negative.

\subsection{Interpretation and Overview of The Soft-Thresholding Scheme}

With  the local detection statistics $W_{k,n}$'s in (\ref{eq:W_n2}),  the soft-thresholding scheme in (\ref{eqn:soft}) can be used to monitor  $K$ data streams with possible local positive or negative mean shifts. It is natural to ask why the the soft-thresholding scheme in (\ref{eqn:soft}) works? Besides the motivation in the previous section, below we also provide a semi-Bayesian interpretation.

At a given time $n,$ let $Z_{k}$ be the indicator whether the distribution of the $k$-th local data stream changes for $k=1, \ldots, K.$ Assume that each local data stream has a prior probability $\pi$ getting affected by the event, and assume that $Z_1, \ldots, Z_{K}$ are iid with probability mass function $\Prob(Z_{k} = 1) = \pi = 1- \Prob(Z_{k} = 0).$ Treat $Z_{k}$'s as the hidden states, and recall that $W_{k,n}$ represents the evidence of possible change (in logarithm scale) and is applicable only when $Z_{k} = 1$ (since $Z_{k} = 0$ implies that there is  no change at the $k$-th data stream). Then when testing $H_0: Z_1 = \ldots = Z_{K} = 0$ (no change),  the  log-likelihood ratio (LLR) statistic of the hidden state $Z_{k}$'s and the observed data $X_{k,n}$'s  is
\begin{eqnarray*}
LLR(n) &=& \sum_{k=1}^{K} \{ Z_{k} (\log \pi + W_{k,n}) + (1-Z_{k}) \log (1-\pi) \}  - \sum_{k=1}^{K} \log (1-\pi) \\
&=& \sum_{k=1}^{K} Z_{k} \{W_{k,n}  - \log((1-\pi) /\pi) \}
\end{eqnarray*}
Since the $Z_{k}$'s are unobservable, it is natural to maximize $LLR(n)$ over $Z_{1}, \ldots, Z_{K} \in \{0,1\}.$ Hence, the maximum likelihood
estimator of the $Z_{k}$'s is that
\[
\hat Z_{k} = \left\{
               \begin{array}{ll}
                 1, & \hbox{if $W_{k,n} \ge \log((1-\pi)/\pi)$} \\
                 0, & \hbox{otherwise }
               \end{array}
             \right. , \quad \mbox{for } k=1,\ldots, K,
\]
and the generalized log-likelihood ratio becomes
\[
\max_{Z_{k}'s} LLR(n) = \sum_{k=1}^{K} \max\{W_{k,n} - \log((1-\pi)/\pi), 0 \},
\]
which is exactly the form of the soft-thresholding scheme $N_{soft}(a)$ in (\ref{eqn:soft}) with $b_{1} = \log((1-\pi)/\pi).$

The above discussion indicates that if we have a prior knowledge that a fraction of $\pi$ sensors will be affected by the event, the censoring
parameter $b_1$ can be chosen as $b_1= \log((1-\pi)/\pi).$ Meanwhile, as in the previous section,  if we want $\eta$ proportion of data streams to send information to the fusion center when no change occurs, one rule of thumb is to choose $b_1 = \log \eta^{-1}.$ It is interesting to see that these two values of $b_1$ are very close when $\eta = \pi$ is small. In practice, the true value of $\pi$ is often unknown, but one may have a preferred $\eta$ value. Hence, in our simulations below we will choose the censoring parameter $b_1=\log \eta^{-1}.$

\medskip
Note that the proposed soft-thresholding scheme $N_{soft}(a)$ in (\ref{eqn:soft}) can be easily implemented in the censoring sensor network context by parallel computing the $K$ local detection statistics $W_{k,n}$'s  recursively through (\ref{eq:W_n2})-(\ref{eq:STn2}) at the local sensor levels. To be more specific, we can use the following $6 K$ registers to adaptively store all past information at each time step after observing new data: $(S_{k}^{(j)}, T_{k}^{(j)},  W_{k}^{(j)})$ for $j=1,2$ and $k=1,2\cdots, K.$ At any given time step $n,$ we can first update the $4 K$ registers in $(S_{k}^{(j)}, T_{k}^{(j)})$ using the past data and compute the $2K$ estimates $\hat \mu_{k}^{(j)}$ of the post-change means $\mu_{k}$'s.
Then after we observe new observations, $(X_{1, n}, \cdots, X_{K, n}),$ we only need to update the $2K$ registers  $W_{k}^{(j)}$'s and compute the values of $K$ local detection statistics $W_{k}$'s, which allows us to easily compute the global monitoring statistic $G.$ Including the $3K$ intermediate variables $(\hat \mu_{k}^{(j)}, W_{k})$ and the global monitoring statistic $G,$ the proposed scheme only needs $9K+1$ registers to adaptively store all relevant information and involves $O(K)$ computations at any given time step $n.$ Moreover, our proposed scheme can be implemented in the context of censoring sensor networks in the previous section where most computations are done at the remote sensors and the communication cost and the computational burden at the fusion center are marginal. Hence,  our proposed scheme is scalable and can be easily implemented to online monitor large-scale data streams over a long time period.

An overview of our proposed $N_{soft}$ scheme is illustrated in the following algorithm:

\begin{center}
\line(1,0){220}
\end{center}
{\bf Algorithm: Implementation of $N_{soft}$ in (\ref{eqn:soft})}

{\bf Initial parameters:} $\rho$, $s$, $t$, and $b_{1}$ for $k=1,\cdots, K.$

{\bf Set:} A terminal threshold $a.$

{\bf Algorithm:}

\qquad {\bf initialize} $n=0,$ and set all initial observations $X_{k} = 0$ and all $8K$ initial registers $S_k^{(j)}=T_k^{(j)}=\mu_k^{(j)}=W_k^{(j)}=0$, for $k=1,\dots,K$ and $j=1,2.$

\qquad {\bf While } the scheme $N_{soft}$  has not raised an alarm

\qquad {\bf do}\quad  1. Update $4K$ registers $(S_{k}^{(j)}, T_{k}^{(j)})$ via (\ref{eq:STn2}).

\qquad \qquad 2. Compute the $2K$ intermediate variables $\hat\mu_k^{(j)}$ from (\ref{eqn:muhat2}) which are the estimates of

\qquad \qquad \quad the post-change means.

\qquad\qquad 3. Input new observations from all $K$ data streams, denoted by $(X_{1}, \cdots, X_{K}).$

\qquad\qquad 4. For $k=1,\ldots,p,$ recompute the local monitoring statistics $W_{k}^{(j)}$'s in (\ref{eq:W_n3}) and
$W_{k}$ in (\ref{eq:W_n2}).

\qquad\qquad 5.  Compute the global monitoring statistics $$G=\sum_{k=1}^{K} \max(W_{k}- b_{1},0)$$

\qquad {\bf if}\quad $G \ge a$ {\bf terminate:} Raising an alarm  at time $n$ and declaring that a change has occurred;

\qquad {\bf end the while loop}
\begin{center}
\line(1,0){220}
\end{center}

%\begin{enumerate}
%\item[1.] Input $\rho$, $s$, $t$, and $d.$
%\item[2.] Initialize $S_k=T_k=\theta_k=W_k=0$, for $k=1,\dots,p$
%\item[3.]
%\item[4.] Compute $\hat\theta_k=\max\left(\rho, \dfrac{s+S_{k}}{t+T_{k}}\right)$, for $k=1,\dots,K$
%\item[5.] Generate a new observation from each data stream $\{X_{k}\}_{k=1}^K$
%\item[6.] Recompute $\{W_{k}\}_{k=1}^K$, where $$
%W_k = \max\left(W_k+\log\left(\dfrac{g_{\hat\theta_k}(X_k)}{f_{\theta_0}(X_k)}\right), 0\right)$$
%\item[7.] Summarize information from $K$ data streams $$G=\sum_{k=1}^K \max(W_{k}-b_k,0)$$
%\item[8.] Repeat steps 3-7 while $G<d$
%\end{enumerate}

\subsection{Simulation Results} \label{sec:results}

In this section, we report the numerical simulation results of the soft-thresholding scheme $N_{soft}(a)$ in (\ref{eqn:soft}).
For the purpose of comparison, we follow Xie and Siegmund \cite{xie2013sequential} to assume that there  are $K=100$ independent normal data streams. For each $k=1, \cdots, K,$ the data $X_{k,n}$'s of the $k$-th data stream are iid $N(0,1)$ before the change, but are iid $N(1,1)$  after  the $k$-th data stream is affected by the occurring event.

In our simulations, we consider six schemes: two of them are the Xie and Siegmund schemes $T_{XS}(a, p_0)$ in (\ref{eqn_XS}) with $p_0 = 1$ and $0.1;$ and  the remaining four schemes are our proposed soft-thresholding schemes $N_{soft}(a)$ in (\ref{eqn:soft}) with four different thresholding parameters: $b_1=0, 0.5, \log(10), \log(100).$ As in the previous section,
the three non-zero $b_1$ values imply that on average at most $\exp(- b_{1}) \approx 60.1\%, 10\%$ and $1\%$ out of $100$ local data streams produce significant $W_{k,n}$'s values to the global monitoring statistic $G_{n}$ when there are no changes. When computing the local detection statistics $W_{k,n}$'s in (\ref{eq:W_n2}), we set $\rho = 0.25, t=4$ and $s= 1$ as in Lorden and Pollak \cite{lorden2008sequential}.

For each of these six schemes $T(a),$ we first numerically search the threshold $a$ to satisfy the  global false alarm constraint $\gamma$ in (\ref{eq000002}). Two different values of $\gamma$ are considered. One is $\gamma = 5000,$
so that we can compare with those results from Xie and Siegmund \cite{xie2013sequential}. The other is $\gamma = 5 \times 10^4$ to
see the effect of false alarm constraint $\gamma$ on the detection delays of our proposed schemes. Note that we are unable to numerically find the global threshold $a$ of the Xie and Siegmund scheme for the case of $\gamma = 5 \times 10^4$ in a reasonable time, and thus we will only report the performance of our proposed schemes.  Next, for the detection delays of $T(a),$ we consider various post-change hypotheses, and for each post-change hypothesis, we simulate the $\E(T(a))$ when the event occurs at time $\nu=1,$ and use this as an estimate of the detection delay ${\bf D}(T(a)).$  All simulated values are based on $2500$ Monte Carlo runs.
%demonstrate the scalability of our proposed schemes. In that case, when we simulate the schemes in Xie and Siegmund,

\begin{table}[t]
  \centering
  \caption{A comparison of detection delays when the change is instantaneous  and the post-change mean $\mu_{k}=1$ if affected. The smallest and largest standard errors of the schemes are also reported under each post-change hypothesis based on $2500$ repetitions in Monte Carlo simulations.}\label{table3}

%\bigskip

\begin{tabular}{|c||c|c|c|c|c|c|c|c|c|c|}
\hline
$\gamma$ & &\multicolumn{9}{|c|}{\# local data streams affected}\\
\cline{3-11}
&&1&3&5&8&10&20&30&50&100\\
\hline
\hline
 &Smallest standard error&0.19& 0.08 &0.06 &0.04 &0.03 &0.02   & 0.01  &0.01 &0.00\\
 &Largest standard error&0.40 & 0.14 &0.08 &0.05 &0.04
 &0.03  &0.02   &0.02  &0.01\\
\hline
\hline
& \multicolumn{10}{c|}{Xie and Siegmund's schemes $T_{XS}(a,p_0)$ in (\ref{eqn_XS})}\\
\cline{2-11}
& $T_{XS}(a=53.5, p_0=1)$   &52.4 & 18.3 & 11.1 & 7.1 & 5.7 & 2.9 & 2.0 & 1.2 & 1.0 \\
& $T_{XS}(a=19.5, p_0=0.1)$ &31.1& 13.4 & 9.2 &6.7  & 5.7 & 3.5 & 2.5 & 1.8 & 1.0 \\
\cline{2-11}
$5000$ &\multicolumn{10}{c|}{Soft-thresholding Schemes $N_{soft}(a)$ in (\ref{eqn:soft})}\\
\cline{2-11}
&$N_{soft}(a=127.86, b_1=0)$&75.0& 35.4& 25.2&18.5 & 16.0&10.3 & 8.1&  6.1 &4.1 \\
&$N_{soft}(a= 84.91, b_1=0.5)$&72.1 &33.9 &24.1&17.7  & 15.3& 10.0  &  7.9& 6.0  &  4.2   \\
&$N_{soft}(a=24.01, b_1=\log(10))$&45.8& 22.0&16.4&12.8 &11.5& 8.5& 7.3& 6.1&  5.0 \\
&$N_{soft}(a=7.88, b_1=\log(100))$&29.0&17.2&14.2 &12.0 & 11.2 & 9.2  & 8.3 & 7.3& 6.4   \\
\cline{2-11}
\hline
\hline
&\multicolumn{10}{c|}{Soft-thresholding Schemes $N_{soft}(a)$ in (\ref{eqn:soft})}\\
\cline{2-11}
&$N_{soft}(a= 136.07, b_1=0)$&89.0& 39.9& 27.9&20.2 & 17.4&11.1 & 8.7&  6.5 &4.4  \\
&$N_{soft}(a= 92.79, b_1=0.5)$&85.7 &38.2 &26.8&19.4  & 16.7& 10.7  &  8.4& 6.3  &  4.4   \\
$5\times 10^4$ &$N_{soft}(a=29.05, b_1=\log(10))$&55.1 & 25.3&18.4&14.1  &12.6& 9.1& 7.8& 6.5&  5.2 \\
&$N_{soft}(a=11.11, b_1=\log(100))$&35.5&19.7&16.0 &13.4 & 12.4 & 10.0  & 8.9 & 7.9& 6.8   \\
 \hline
\end{tabular}
\end{table}

Table \ref{table3} summarizes the detection delays in the scenario when the change is instantaneous if a local data stream is affected.
For the Xie and Siegmund scheme $T_{XS}(a, p_0)$ in  (\ref{eqn_XS}), our simulated detection delay results are slightly different from their reported  results in their paper, possibly because our simulation is based on $2500$ runs instead of $500$ runs in their paper.
Note that the Xie and Siegmund schemes $T_{XS}(a, p_0)$ in  (\ref{eqn_XS}) involve expensive computations, and require the fusion center to have full access to all raw data. Thus it is not surprising that their schemes have smaller detection delays than our proposed soft-thresholding schemes. However, we want to emphasize that the  Xie and Siegmund schemes are not scalable and cannot be implemented in the context of distributed monitoring in censoring sensor networks. Meanwhile, our proposed schemes are suitable to the censoring sensor network contexts, as they can be easily implemented by parallel computing in a recursive manner at the local sensors level and the computational costs between the local sensors and the fusion center will be marginal.

%  the differences are not dramatic, e.g., when $10$ out of $100$ data streams are affected, the smallest detection delay of their schemes is $6.7$ as compared to $11.3$ of our proposed soft-thresholding  schemes.

A more reasonable comparison is to compare the results in Table \ref{table3}  with those in Table \ref{table01} which were conducted under the assumption that the post-change mean of each affected local data stream is $\mu=1.$ When at least $5$ local data streams are affected, the detection delays of $N_{soft}(a=24.01, b_{1}= \log(10))$ in Table \ref{table3} are only $2 \sim 5$ larger than those of $N_{soft}(a= 21.56, b_{k} = 2.3026)$ in Table \ref{table01}. Since the schemes in Table \ref{table3} are able to detect both positive or negative mean shifts, one may be willing to pay the price of slightly larger detection delays at the given post-change mean $\mu=1$ so as to effectively detect other local mean shifts, especially the negative shifts. In addition, it is interesting to see from Table \ref{table3} that as the false alarm constraint $\gamma$ increases from $5000$ to $5 \times 10^4,$ the global threshold $a$ of our proposed soft-thresholding schemes $N_{soft}(a, b_1)$ increases moderately for any given censoring parameters $b_1,$ but the detection delays of our proposed soft-thresholding schemes increase only marginally when at least $5$ data streams are affected.

All simulations were done on a Windows 8 Laptop with Intel i7-4700MQ CPU \@2.40GHz using MATLAB R2013b. For each of these schemes $T(a)$ (i.e., each row of Table \ref{table3}), the most time consuming part was to search for the global threshold $a$ so that $\E^{(\infty)}(T(a)) \approx \gamma.$ When $\gamma = 5000,$ it took about $8$ minutes to find such $a$ from {\it a range of values} for our proposed schemes based on $2500$ Monte Carlo runs (the time is shorter if our initial guess range of $a$ is closer). Meanwhile, for the Xie and Siegmund scheme, for a {\it given} global threshold $a$ around $53.5$ which was provided in their paper, it took about one and a half hour on average to finish one Monte Carlo simulation run in our laptop. If we did not know $a \approx 53.5$ and wanted to try $10$ different values of $a$'s by bisection method based on $2500$ Monte Carlo runs for each $a$,  it would have taken about $10 \times 1.5 \times 2500 = 37500$ computer hours for the case of $\gamma = 5000.$ When $\gamma = 5 \times 10^4,$  it took us about one hour to find the global threshold $a$ for our  proposed schemes, but we are unable to numerically implement the Xie and Siegmund schemes since their computational time will be in days for each Monte carlo run. Once the global threshold $a$ is found, it is straightforward to simulate the detection delays in Table \ref{table3}. When $\gamma = 5000,$ our proposed schemes are at least $10$ times faster than the Xie and Siegmund schemes. For instance, when exactly one data stream is affected, it took $4.94$ seconds to simulate the detection delay of our proposed schemes, whereas it took $41.02$ seconds to simulate those of the Xie and Siegmund schemes. Hence, as compared to the Xie and Siegmund schemes, the computational advantage of our proposed schemes is evident.

\section{Proof of Theorems \ref{theorem1} and \ref{theorem2}}

This section is devoted to prove Theorems \ref{theorem1} and \ref{theorem2}.

{\bf Proof of Theorem \ref{theorem1}.} Intuitively, only those affected sensors provide information to detect the occurring events,
and the quickest possible way to detect the  occurring event is when the event affects the sensors instantaneously. More rigorously, if we define
\begin{eqnarray} \label{newdelta}
\delta_{k}^{*} = \left\{
                   \begin{array}{ll}
                     0, & \hbox{if $\delta_{k}$ is finite} \\
                     \infty, & \hbox{if $\delta_{k}=\infty$}
                   \end{array}
                 \right. ,
\end{eqnarray}
then for any given scheme $T(\gamma),$
\[
\overline{\E}_{\delta_1,  \ldots, \delta_{K}}(T(\gamma)) \ge \inf_{\tau} \overline{\E}_{\delta_1^{*},  \ldots, \delta_{K}^{*}}(\tau),
\]
where the  infumum  is taken over all possible schemes $\tau$ satisfying the false alarm constraint $\gamma$ in (\ref{eq000002}).
An alternative and possible better viewpoint is based on
a time-shifting argument in which one imagines that at time $n$
one observes the observations $X_{k, n+\delta_{k}}$ (instead of $X_{k,n}$) when $\delta_{k}$ is finite, and then applies $T(\gamma)$ to the new aligned observations.

Without loss of generality, assume that the first $m$ data streams
are affected abruptly and simultaneously by the event at unknown time $\nu,$ and other data streams are unaffected.
That is, $m$ out of $K$ data streams are affected by the event, and
$\delta_{i}^{*}=0$ for $1 \le i \le m,$ and $=\infty$ for $m+ 1\le i \le K.$ By (\ref{Jinfo}), we have
$$J(\delta_1, \ldots, \delta_{K})=J(\delta_1^{*}, \ldots, \delta_{K}^{*}) = \sum_{i=1}^{m} I(g_{i}, f_{i}).$$
In this case, we face the sequential change detection problem when the distribution of $(X_{1, n}, \cdots, X_{K,n})$ changes from $(f_1, \cdots, f_{m}, f_{m+1}, \cdots, f_{K})$ to $(g_1, \cdots, g_{m}, f_{m+1}, \cdots, f_{K}).$  It is well-known (Lorden \cite{lorden:1971}) that
\[
\inf_{\tau} \overline{\E}_{\delta_1^{*},  \ldots, \delta_{K}^{*}}(\tau) \ge (1+o(1)) \frac{\log \gamma}{\sum_{i=1}^{m} I(g_{i}, f_{i})}.
\]
subject to the false alarm constraint $\gamma$ in (\ref{eq000002}) as $\gamma \rightarrow \infty.$ Combining the above results yields relation (\ref{eqlowerbound}), completing the proof of Theorem \ref{theorem1}. \hspace*{\fill}~\QED

\medskip
{\bf Proof of Theorem \ref{theorem2}.} Let us first focus part (i) on
the properties of the hard-thresholding scheme $N_{hard}(a,b)$  in (\ref{eq000012}) with $b \ge 0$ being the common constant for $b_{k}$'s in (\ref{eq000010})-(\ref{eq000011}).

To prove (\ref{eq000006}), note that   $N_{hard}(a,b)$ in (\ref{eq000012})
is increasing as a function of $b \ge 0,$ and when $b=0,$ $N_{hard}(a, b=0)$ reduces to the
``SUM" scheme $T_{\rm sum}(a)$ in (\ref{eq3n03}). Hence, for any $b \ge 0,$ $N_{hard}(a,b) \ge
T_{\rm sum}(a)$ and of course, $\E^{(\infty)}(N_{hard}(a,b)) \ge \E^{(\infty)}(T_{\rm sum}(a)).$
By Theorem  1 of Mei \cite{mei:2009}, the ``SUM" scheme $T_{\rm sum}(a)$  satisfies relation (\ref{eq000006}), and so
are the hard-thresholding schemes $N_{hard}(a,b)$ for all $b \ge 0.$

To prove relation (\ref{eq0011}), it is clear that the worst-case detection delay of $N_{hard}(a,b)$ occurs at the
change-point $\nu=1,$ and thus it suffices to show that
$\E_{\delta_1,\ldots,\delta_{K}}^{(\nu=1)}(N_{hard}(a,b))$ satisfies
(\ref{eq0011}). Without loss of generality, we assume that only the first $m$ data steams are affected and no
other data streams are affected. To   simplify our notation below, denote $\delta_{\max} = \max_{1 \le i \le m} \delta_{i}.$
%Since $N_{hard}(a,b)$ is increasing as a function of $b \ge 0,$
It suffices to show that
\begin{eqnarray} \label{eq0018}
\E_{\delta_1,\ldots,\delta_{K}}^{(\nu=1)}(N_{hard}(a,b))  \le \frac{a }{\sum_{k=1}^{m} I(g_{k},
f_{k})} + O(\sqrt{b}) + O(1) + \delta_{\max},
\end{eqnarray}
for all $0 \le b \le b'$ when $b'$ and $(a-b')$ go to $\infty.$

The essential idea in the proof of (\ref{eq0018}) is to compare $N_{hard}(a,b)$ with  new stopping times  that are only based on those affected $m$ data streams.  Define a stopping time that is in the form of the one-sided sequential probability ratio test (SPRT):
\begin{eqnarray} \label{newoneSPRT}
\tau(a,b) = \mbox{first $n$ such that } \sum_{i=1}^{n} \sum_{k=1}^{m} \log
\frac{g_{k}(X_{k,i})}{f_{k}(X_{k,i})} \ge a\  \mbox{ and }\
\sum_{i=1}^{n} \log \frac{g_{k}(X_{k,i})}{f_{k}(X_{k,i})} \ge
\rho_{k} b \ \mbox{ for all $1 \le k \le m$},
\end{eqnarray}
where the weights $\rho_{k}$'s are  defined in (\ref{eq000011}), and let $\hat \tau_{\delta}(a,b)$ be the new stopping time that applies $\tau(a,b)$ to the new observations after time $\delta_{\max}.$

Now whenever  $\hat \tau_{\delta}(a,b)$ stops at time $n_0+\delta_{\max},$ we know that $\tau(a,b)$ stops after applying it to $n_0$ observations $(X_{k, \delta_{\max}+1}, \cdots, X_{k, \delta_{\max} + n_0})$ for each $k.$ By the definition of the local CUSUM statistics in (\ref{eq3n01}), we have
\[
W_{k,n_0+\delta_{\max}} \ge \sum_{i=\delta_{\max}+1}^{\delta_{\max}+n_0} \log \frac{g_{k}(X_{k,i})}{f_{k}(X_{k,i})} \ge \rho_{k} b
\]
for all $1 \le k \le m.$ Hence,
\[
\sum_{k=1}^{K} W_{k,n_0+\delta_{\max}} {\bf 1}\{W_{k,n_0+\delta_{\max}} \ge \rho_{k} b \}  \ge \sum_{k=1}^{m} \sum_{i=\delta_{\max}+1}^{\delta_{\max}+n_0} \log \frac{g_{k}(X_{k,i})}{f_{k}(X_{k,i})} \ge a,
\]
where the last relation is from the definition of $\tau(a,b).$ This implies that the scheme $N_{hard}(a,b)$ must stop at time $n_0+\delta_{\max},$ and possibly earlier. Thus
\begin{eqnarray*}
\E_{\delta_1, \ldots, \delta_{K}}^{(\nu=1)}(N_{hard}(a,b)) &\le& \E^{(\nu=1)}_{\delta_1, \ldots, \delta_{K}}(\hat \tau_{\delta}(a,b)) = \delta_{\max} + \E^{(\nu=1)}_{\delta_1^{*}, \ldots, \delta_{K}^{*}} (\tau(a,b)),
\end{eqnarray*}
where $\delta_{k}^{*}$ is  the binary version of $\delta_{k}$'s defined in (\ref{newdelta}). To simplify the notation, denote by $\E^{(1)}$ the expectation
when the change occurs at time $\nu=1$ and the event affects the first $m$ data streams immediately but does not affect the other remaining $K-m$ data streams. So it
suffices to show that the stopping time  $\tau(a,b)$ in  (\ref{newoneSPRT})  satisfies
\begin{eqnarray} \label{eqnn0020}
\E^{(1)}(\tau(a,b))  \le \frac{a }{\sum_{k=1}^{m} I(g_{k}, f_{k})} + O(\sqrt{b}) + O(1).
\end{eqnarray}

To prove (\ref{eqnn0020}),  for $1 \le k \le m,$ let
\begin{eqnarray*}
M_{k} &=& \inf \Big\{ n \ge 1:  \sum_{i=1}^{n} \log
\frac{g_{k}(X_{k,i})}{f_{k}(X_{k,i})} \ge \rho_{k} b  \Big\}, \\
\tau_{k}(M_{k}) &=& \sup \Big\{ n \ge 1:
\sum_{i=M_{k}+1}^{M_{k}+n} \log \frac{g_{k}(X_{k, i})}{f_{k}(X_{k,
i})} \le 0 \Big\} \\
\hat M &=& \max_{1 \le k \le m} \Big(M_{k} + \tau_{k}(M_{k}) + 1
\Big)
\\
t(\hat M) &=& \inf \Big\{n \ge 1: \sum_{i=\hat M+1}^{\hat M+n}
\Big( \sum_{k=1}^{m} \log \frac{g_{k}(X_{k, i})}{f_{k}(X_{k, i})}
\Big) \ge a - (\sum_{k=1}^{m} \rho_{k}) b \Big\}.
\end{eqnarray*}
In the definition of $t(\hat M),$ the assumption of $0 \le b \le b'$ is used to  make sure that
the threshold
\[
a - (\sum_{k=1}^{m} \rho_{k}) b \ge a -b \ge a - b'
\]
goes to $\infty$ as $a \rightarrow \infty,$
since  $\sum_{k=1}^{m} \rho_{k} \le \sum_{k=1}^{K} \rho_{k} = 1$ and $(a - b')$ is assumed to go to $\infty.$
Combining these definitions with those of $\tau(a,b)$ in  (\ref{newoneSPRT}) yields that
\begin{eqnarray*}
\tau(a,b) &\le& \hat M + t(\hat M) = \max_{1 \le k \le m}
\Big(M_{k} + \tau_{k}(M_{k})+1\Big) + t(\hat M) \cr &\le&
\sum_{k=1}^{m} \tau_{k}(M_{k}) + 1  +t(\hat M) + \max_{1 \le k \le m} M_{k}.
\end{eqnarray*}
Hence, relation (\ref{eqnn0020}) holds if we can establish the following three relations:
\begin{eqnarray}
\E^{(1)}\Big(\tau_{k}(M_{k})\Big) &=& O(1) \qquad \mbox{for all $1 \le k \le m$};  \label{eqqn1} \\
\E^{(1)}\Big(t(\hat M)\Big) &\le& \frac{a}{\sum_{k=1}^{m} I(g_{k}, f_{k})} - \frac{b}{\sum_{k=1}^{K}
I(g_{k}, f_{k})} + O(1); \label{eqqn2}\\
\E^{(1)}\Big(\max_{1 \le k \le m} M_{k}\Big) &\le& \frac{b}{\sum_{k=1}^{K}I(g_{k},f_{k})}+O(\sqrt{b}) + O(1). \label{eqqn3}
\end{eqnarray}
Relation (\ref{eqqn1}) is well-known in renewal theory, e.g., Theorem D in Kiefer and Sacks
\cite{kiefer:1963}, since $\log \big(g_{k}(X)/f_{k}(X)\big)$ has positive mean and
finite variance under $\E^{(1)}$ by our assumptions in (\ref{eq000001}) and (\ref{eq000001b}).

For relation (\ref{eqqn2}), by the definition of
$\rho_{k}$ in (\ref{eq000011}), we have
\[
\frac{\sum_{k=1}^{m} \rho_{k}}{\sum_{k=1}^{m} I(g_{k}, f_{k})} = \frac{1}{\sum_{k=1}^{K} I(g_{k}, f_{k})}.
\]
Since $t(\hat M)$ is the stopping time when a random walk exceed the bound $a- (\sum_{k=1}^{m}
\rho_{k})b,$  the application of standard renewal theory  yields that
\begin{eqnarray*}
\E^{(1)}(t(\hat M)) &=& \frac{a- (\sum_{k=1}^{m}
\rho_{k})b}{\sum_{k=1}^{m} I(g_{k}, f_{k})} + O(1) \cr &=&
\frac{a}{\sum_{k=1}^{m} I(g_{k}, f_{k})} - \frac{b}{\sum_{k=1}^{K}
I(g_{k}, f_{k})} + O(1),
\end{eqnarray*}
as the threshold $a- (\sum_{k=1}^{m} \rho_{k})b$  goes to $\infty,$ see, for example, Siegmund \cite[Ch. VIII]{siegmund:1985}. Thus relation (\ref{eqqn2}) holds.

The proof of relation (\ref{eqqn3}) is a little more complicated, but it can be done
along the same line as that in Mei \cite{mei:2005}.  The key fact is that the choice of $b_{k}=\rho_{k} b$'s in (\ref{eq000010})-(\ref{eq000011}) makes sure that the stopping times $M_{k}$'s have roughly the same mean under $\Prob^{(1)}.$ Specifically,
by renewal theory and the assumptions of $(f_{k}, g_{k})$ in (\ref{eq000001}) and (\ref{eq000001b}), under $\Prob^{(1)},$
\[
\E^{(1)}(M_{k}) = \frac{\rho_{k} b}{ I(g_{k}, f_{k})} + O(1) =
\frac{b}{ \sum_{k=1}^{K} I(g_{k}, f_{k})} + O(1)
\]
and $\mbox{Var}^{(1)}(M_{k}) = O(b),$ as $b \rightarrow \infty,$
see Siegmund \cite[p. 171]{siegmund:1985}. Thus
\begin{eqnarray*}
\Big(\E^{(1)} \big| M_{k}-\frac{b}{\sum_{k=1}^{K} I(g_{k},
f_{k})}\big|\Big)^{2} &\le& \E^{(1)} \Big(
M_{k}-\frac{b}{\sum_{k=1}^{K} I(g_{k}, f_{k})}\Big)^{2} \\
&=& \mbox{Var}^{(1)}(M_{k}) + \Big(\E^{(1)}
M_{k}-\frac{b}{\sum_{k=1}^{K} I(g_{k}, f_{k})}\Big)^{2}
\\
&=& O(b)
\end{eqnarray*}
as $b \rightarrow \infty.$ Hence, for each $k=1, \cdots, K,$ there exist two constants $C_{1k} > 0$ and $C_{2k} > 0$ so that for all $b \ge 0,$
\[
\big| M_{k}-\frac{b}{\sum_{k=1}^{K} I(g_{k},
f_{k})}\big| \le \max(C_{1k}, C_{2k} \sqrt{b}).
\]
Therefore,
\begin{eqnarray*}
\E^{(1)} \Big(\max_{1 \le k \le m} M_{k} \Big) &=&
\frac{b}{\sum_{k=1}^{K} I(g_{k}, f_{k})} + \E^{(1)} \max_{1 \le k
\le m} \Big(M_{k}-\frac{b}{\sum_{k=1}^{K} I(g_{k}, f_{k})}\Big)
\cr &\le& \frac{b}{\sum_{k=1}^{K} I(g_{k}, f_{k})} +
\sum_{k=1}^{m} \E^{(1)} \Big|M_{k}-\frac{b}{\sum_{k=1}^{K}
I(g_{k}, f_{k})}\Big| \cr\cr &\le& \frac{b}{\sum_{k=1}^{K} I(g_{k},
f_{k})} + \sum_{k=1}^{m} \max(C_{1k}, C_{2k} \sqrt{b}) \cr
&\le& \frac{b}{\sum_{k=1}^{K} I(g_{k}, f_{k})} + C (\sqrt{b}+1),
\end{eqnarray*}
where the constant $C = \sum_{k=1}^{K} \max(C_{1k}, C_{2k})$ does not depend on $b.$
This proves relation (\ref{eqqn3}). Therefore, relations (\ref{eqqn1})-(\ref{eqqn3}) hold, and thus relation (\ref{eq0011}) holds for the hard-thresholding scheme $N_{hard}(a,b)$ in (\ref{eq000012}).   The proof for the soft-thresholding scheme $N_{soft}(a,b)$ in (\ref{eq000012b}) is identical and thus omitted.

Now let us provide a sketch of the proof for part (iii) of Theorem \ref{theorem2} on the order-thresholding scheme $N_{order, r}(a)$ in (\ref{rankCUSUM}) and the combined thresholding scheme $N_{comb, r}(a,b)$ in (\ref{rankCUSUM1}).
Since $N_{order, r}(a)$ is a special case of $N_{comb, r}(a,b)$ with $b=0,$ it suffices to prove the theorem for $N_{comb, r}(a,b)$ in (\ref{rankCUSUM1}) with $b \ge 0.$
Clearly relation   (\ref{eq000006}) also holds for $N_{comb, r}(a,b)$ for any $b \ge 0,$ because the
``SUM" scheme $T_{\rm sum}(a)$ again provides the lower bound for $N_{comb, r}(a,b).$
%The properties of false alarms are straightforward since and thus

It remains to show that relation (\ref{eq0011}) holds for $N_{comb, r}(a,b)$ with $b \ge 0$ in the scenario when the occurring event affects at most $r$ data streams, i.e., when $\sum_{k=1}^{K} I\{ \delta_{k} < \infty\} \le r.$
 Without loss of generality, assume that the affected data streams are just the first $m$ data streams with $m \le r.$
Recall that $U_{k,n} = W_{k,n} I\{W_{k,n} \ge \rho_{k} b\},$ and we order
the $U_{k,n}$'s as $U_{(1), n} \ge \ldots \ge U_{(K),n},$
and  $N_{comb, r}(a,b)$ stops if $\sum_{k=1}^{r} U_{(k),n} \ge a.$
Note that if $ m \le r,$
\begin{eqnarray*}
\sum_{k=1}^{r} U_{(k),n} &\ge& \sum_{k=1}^{r} U_{k,n} \ge \sum_{k=1}^{m} U_{k,n},
\end{eqnarray*}
since $U_{k,n} \ge 0.$ Thus, if at some time $n_0$
we have $W_{k, n_0} \ge \rho_{k} b$ and $\sum_{k=1}^{m} W_{k,n_0} \ge a$ for $1 \le k \le m$ (i.e., for the first $m$ data streams),
then  $N_{comb, r}(a,b)$ will also stop at time $n_0$ and possibly earlier. Hence, whenever $m \le r,$
the stopping time $\tau(a,b)$ in (\ref{newoneSPRT}) also provides an upper bound
on the detection delay of  $N_{comb, r}(a,b).$ Thus the proposed combined thresholding scheme $N_{comb, r}(a,b)$ in (\ref{rankCUSUM1})
satisfies relation (\ref{eq0011}) whenever
the occurring event affects at most $r$ data streams.
This completes the proof of the theorem. \hspace*{\fill}~\QED


\begin{thebibliography}{1}

\bibitem{appadwedula:2005}
{\sc Appadwedula,} S., {\sc Veeravalli,} V. V., and {\sc Jones,}
D.  L. (2005). Energy-efficient detection in sensor networks. {\em
IEEE J. Sel. Areas Commun.}, {\em 23}, 693--702.


\bibitem{bass:1993}
{\sc Basseville,} M. and {\sc Nikiforov,} I.
V. (1993).
\emph{Detection of Abrupt Changes: Theory and Applications.}
Englewood Cliffs, Prentice-Hall. MR1210954


\bibitem{breiman:2001}
{\sc Breiman,} L. (2001). Statistical modeling: the two cultures. {\em
Statistical Sciences}, {\em 16}, 199--231. MR1874152 

\bibitem{candes:2006}
{\sc Cand\`es}, E. J. (2006). Modern statistical estimation via
oracle inequalities. {\it Acta Numerica}, {\em 15}, 257--325.
MR2269743

\bibitem{donoho:94}
{\sc Donoho}, D. L. and {\sc Johnstone}, I. M. (1994).
Ideal spatial adaptation by wavelet shrinkage. {\it Biometrika},
\textbf{81}, 425--455. MR1311089

\bibitem{durrett:1996}
{\sc Durrett}, R. (1996). \emph{Probability: Theory and Examples}.
Second edition. Duxbury Press, Belmont, CA. MR1609153


\bibitem{Fan:98}
{\sc Fan}, J. and {\sc Lin}, S. K. (1998). Test of significance when data are curves.
{\it Journal of American Statistical Association}, \textbf{93}, 1007--1021. MR1649196

\bibitem{Fuh:2015}
{\sc Fuh,} C.D. and {\sc Mei,} Y. (2015). Quickest change detection and
Kullback-Leibler divergence for two-state hidden Markov models.
{\em IEEE Trans. Signal Processing}, \textbf{63}, 4866--4878.


\bibitem{scan}
{\sc Glaz}, J., {\sc Naus}, J. and {\sc Wallenstein}, S. (2001). \emph{Scan Statistics}.
Springer-Verlag, New York. 
MR1869112 



\bibitem{Gordon:1994}
{\sc Gordon,} L. and {\sc Pollak,} M. (1994). An efficient sequential nonparametric
scheme for detecting a change of distribution. \emph{Ann. Statist.} \textbf{22} 763--804. MR1292540

\bibitem{kiefer:1963}
{\sc Kiefer,} J. and {\sc Sacks,} J. (1963). Asymptotically
optimum sequential inference and design. \emph{Ann. Math. Statist.} \textbf{34} 705--750. MR0150907

\bibitem{kulldorff:2001}
{\sc Kulldorff,} M. (2001). Prospective Time-Periodic Geographic
Disease Surveillance Using a Scan Statistic, \emph{J.
R. Stat. Soc. Ser. A} \textbf{164} 61--72. 
MR1819022

\bibitem{lai:1995}
{\sc Lai,} T. L. (1995). Sequential change-point detection in
quality control and dynamical systems (with discussion). {\it J.
R. Stat. Soc. Ser. B Stat. Methodol.} \textbf{57} 613--658.
MR1354072 


\bibitem{lai:2001}
{\sc Lai,} T. L. (2001). Sequential analysis: some classical
problems and new challenges. \emph{Statist. Sinica} \textbf{11}
303--408. MR1844531

\bibitem{levy:2009}
{\sc L{\'e}vy-Leduc,} C. and {\sc Roueff,} F. (2009). Detection and localization of change-points in
high-dimensional network traffic data. \emph{Ann. Appl. Stat.} \textbf{3} 637--662. 
MR2750676


\bibitem{liu:2015}
{\sc Liu,} K., {\sc Mei,} Y.,  and {\sc Shi,} J.  (2015). An adaptive
sampling  strategy for online high-dimensional process monitoring. \emph{Technometrics} \textbf{57}
305--319. MR3384946

\bibitem{lorden:1971}
{\sc Lorden,} G. (1971). Procedures for reacting to a change in
distribution. \emph{Ann. Math. Statist.} \textbf{42} 1897--1908.
MR0309251


\bibitem{lorden2008sequential}
{\sc Lorden,} G. and {\sc Pollak,} M. (2008).
Sequential change-point detection procedures that are nearly optimal
and computationally simple. \emph{Sequential Analysis} \textbf{27} 476-512. 
MR2460209 


\bibitem{mei:2005}
{\sc Mei,} Y. (2005).  Information bounds and quickest change
detection in decentralized decision systems.  \emph{IEEE Trans.
Inform. Theory} \textbf{51} 2669--2681. MR2246385



\bibitem{mei:2009}
{\sc Mei,} Y. (2010). Efficient scalable schemes for monitoring
a large number of data streams.  \emph{Biometrika} \textbf{97.2} 419-433. MR2650748


\bibitem{Montgomery:1991}
{\sc Montgomery}, D. C. (1991). \emph{Introduction to Statistical
Quality Control} (2nd edition). Wiley, New York.



\bibitem{moustakides:1986}
{\sc Moustakides,} G. V. (1986). Optimal stopping times for
detecting changes in distributions. \emph{Ann. Statist.}
\textbf{14} 1379--1387. MR0868306


\bibitem{neyman:37}
{\sc Neyman,} J. (1937). Smooth test for goodness-of-fit.
\emph{Skand. Aktuarietidskr.} \textbf{20} 149--199.



\bibitem{page:1954}
{\sc Page,} E. S. (1954). Continuous inspection schemes.
\emph{Biometrika} \textbf{41} 100--115. MR0088850


\bibitem{pollak:1985}
{\sc Pollak,} M. (1985). Optimal detection of a change in
distribution. {\em Ann. Statist.} \textbf{13} 206--227. 
MR0773162 


\bibitem{pollak:1987}
{\sc Pollak,} M. (1987). Average run lengths of an optimal method
of detecting a change in distribution. {\em Ann. Statist.}
\textbf{15} 749--779. MR0888438


\bibitem{poor2009quickest}
{\sc Poor,} H. V. and {\sc Hadjiliadis,} O. (2009). \emph{Quickest Detection}.
Cambridge Univ. Press, New York, 2009. 
MR2482527 



\bibitem{rago:1996}
{\sc Rago,} C., {\sc Willett,} P., and {\sc Bar-Shalom,} Y.
(1996). Censoring sensors: A low-communication-rate scheme for
distributed detection. {\em IEEE Trans. Aerosp. Electon. Syst.},
{\em 32}, 554--568.

\bibitem{roberts:1966}
{\sc Roberts,} S. W. (1966). A comparison of some control chart
procedures. \emph{Technometrics} \textbf{8} 411--430. MR0196887


\bibitem{shewhart:1931}
{\sc Shewhart,} W. A. (1931). {\em Economic Control of Quality of
Manufactured Product}. D Van Norstrand, New York. Preprinted by
ASQC Quality Press, Wisconsin, 1980.


\bibitem{shiryayev:1963}
{\sc Shiryaev,} A. N. (1963). On optimum methods in quickest
detection problems. \emph{Theory Probab. Appl.} \textbf{8} 22--46.


\bibitem{siegmund:1985}
{\sc Siegmund,} D. (1985): \emph{Sequential Analysis: Tests and
Confidence Intervals}. Springer, New York. MR0799155

\bibitem{tartakovsky2014sequential}
{\sc Tartakovsky,} A.,  {\sc Nikiforov,} I., and {\sc Basseville}, M. (2015).
\emph{Sequential Analysis: Hypothesis Testing and Changepoint Detection}.
 Monographs on Statistics and Applied Probability, 136. CRC Press, Boca Raton, FL. MR3241619


\bibitem{tartakovsky:2006}
{\sc Tartakovsky,} A. G., {\sc Rozovskiia,} B. L., {\sc Blazeka,}
R. B. and {\sc Kim,} H. (2006). Detection of intrusions in
information systems by sequential change-point methods (with
discussions). {\em Statistical Methodology} \textbf{3} 252--340.
MR2240956


\bibitem{tartakovsky:2004}
{\sc Tartakovsky,} A. G. and {\sc Veeravalli,} V. V. (2004).
Change-point Detection in Multichannel and Distributed Systems.
\emph{Applied Sequential Methodologies}, 339--370, Statist.
Textbooks Monogr., 173, Dekker, New York. 
MR2159163


\bibitem{tay:2007}
{\sc Tay,} W. P., {\sc Tsitsiklis,} J. N. and {\sc Win,} M. Z.
(2007). Asymptotic performance of a censoring sensor network. {\em
IEEE Trans. Inform. Theory} \textbf{53} 4191--4209.
MR2446562

\bibitem{xie:2013a}
{\sc Xie,} Y., {\sc Huang,} J., and {\sc Willett,} R. (2013).
Changepoint detection for high-dimensional time series with missing data.
{\em IEEE Journal of Selected Topics in Signal Processing}, \textbf{7}, 12--27.

\bibitem{xie2013sequential}
{\sc Xie,} Y. and {\sc Siegmund,} D. (2013).
Sequential multi-sensor change-point detection.  \emph{Ann. Stat.},
\textbf{41} 670--692. MR3099117


\bibitem{vvv:2001}
{\sc Veeravalli,} V. V. (2001). Decentralized quickest change detection. {\em
IEEE Trans. Inform. Theory} \textbf{47} 1657--1665. 
MR1830119 


\end{thebibliography}
\end{document}